\documentclass[fleqn,usenatbib]{mnras}

\usepackage{amssymb}
\usepackage{tabularx}
\usepackage{array}
\usepackage{makecell}
\usepackage{multirow}
\usepackage{newtxtext,newtxmath}
\usepackage{ulem}
\usepackage{xcolor}
\usepackage{threeparttable}
\usepackage[T1]{fontenc}

\DeclareRobustCommand{\VAN}[3]{#2}
\let\VANthebibliography\thebibliography
\def\thebibliography{\DeclareRobustCommand{\VAN}[3]{##3}\VANthebibliography}

\usepackage{graphicx}	% Including figure files
\usepackage{amsmath}	% Advanced maths commands

\title[CSST Emission-line Redshift]{CSST Large Scale Structure Analysis Pipeline: III. Emission-line Redshift Measurement for Slitless Spectra}

\author[J. Sui et al.]{Jipeng Sui,$^{1,2}$
Hu Zou,$^{1,2}$\thanks{E-mail: zouhu@nao.cas.cn}
Xiaohu Yang,$^{3,4}$
Xianzhong Zheng,$^{5,6}$
Run Wen,$^{5,6}$
Yizhou Gu,$^{4}$
Weiyu Ding,$^{1,6,7}$
\newauthor Lu Feng,$^{1}$
Hong Guo,$^{8}$
Wei-Jian Guo,$^{1}$
Yunkun Han,$^{9}$
Yipeng Jing,$^{3,4}$
Cheng Li,$^{10}$
Wenxiong Li,$^{1}$
Shufei Liu,$^{1,2}$
\newauthor Zhixia Shen,$^{1}$
Gaurav Singh,$^{1}$
Jiali Wang,$^{1}$
Peng Wei,$^{11}$
Yunao Xiao,$^{1,2}$
Suijian Xue,$^{1}$
Hu Zhan,$^{12,13}$
\newauthor Pengjie Zhang,$^{3,4}$
Gongbo Zhao$^{1,2}$
\\
$^{1}$Key Laboratory of Optical Astronomy, National Astronomical Observatories, Chinese Academy of Sciences, Beijing 100101, China\\
$^{2}$School of Astronomy and Space Science, University of Chinese Academy of Sciences, Beijing 101408, China\\
$^{3}$Department of Astronomy, School of Physics and Astronomy, Shanghai Jiao Tong University, Shanghai 200240, China\\
$^{4}$Tsung-Dao Lee Institute, and Key Laboratory for Particle Physics, Astrophysics and Cosmology, Ministry of Education, Shanghai Jiao Tong University,\\Shanghai 201210, China\\
$^{5}$Purple Mountain Observatory, Chinese Academy of Sciences, 10 Yuanhua Road, Nanjing 210023, China\\
$^{6}$School of Astronomy and Space Science, University of Science and Technology of China, Hefei 230026, China\\
$^{7}$Deep Space Exploration Laboratory / Department of Astronomy, University of Science and Technology of China, Hefei 230026, China\\
$^{8}$Key Laboratory for Research in Galaxies and Cosmology, Shanghai Astronomical Observatory; Nandan Road 80, Shanghai 200030, China\\
$^{9}$Yunnan Observatories, Chinese Academy of Sciences, 396 Yangfangwang, Guandu District, Kunming 650216, China\\
$^{10}$Department of Astronomy, Tsinghua University, Beijing 100084, China,\\
$^{11}$Xinjiang Astronomical Observatory, Chinese Academy of Sciences, Urumqi, Xinjiang 830011, China\\
$^{12}$Key Laboratory of Space Astronomy and Technology, National Astronomical Observatories, Chinese Academy of Sciences, Beijing, 100101, China\\
$^{13}$Kavli Institute for Astronomy and Astrophysics, Peking University, Beijing 100871, China}

\date{Accepted XXX. Received YYY; in original form ZZZ}

\pubyear{}

\begin{document}
\label{firstpage}
\pagerange{\pageref{firstpage}--\pageref{lastpage}}
\maketitle

\begin{abstract}
The China Space Station Telescope (CSST) is a forthcoming space-based optical telescope designed to co-orbit with the Chinese Space Station. With a planned slitless spectroscopic survey spanning a broad wavelength range of $255-1000$\,nm and an average spectral resolution exceeding 200, the CSST holds significant potential for cosmic large-scale structure analysis. In this study, we focus on redshift determinations from slitless spectra through emission line analysis within the CSST framework. Our tailored redshift measurement process involves identifying emission lines in one-dimensional slitless spectra, aligning observed wavelengths with their rest-frame counterparts from prominent galaxy emissions, and calculating wavelength shifts to determine redshifts accurately. To validate our redshift measurement algorithm, we leverage simulated spectra generated by the CSST emulator for slitless spectroscopy. The outcomes demonstrate a remarkable redshift completeness exceeding 95 per cent for emission line galaxies (ELGs), alongside a purity surpassing 85 per cent. The redshift uncertainty remains impressively below than $\sim 0.001$. Notably, when concentrating on galaxies with more than three matched emission lines, the completeness of ELGs and the purity of measurable galaxies can reach 98 per cent and 97 per cent, respectively. Furthermore, we explore the influence of parameters like magnitude, spectral signal-to-noise ratio, and redshift on redshift completeness and purity. The discussion also delves into redshift degeneracies stemming from emission-line matching confusion. Our developed redshift measurement process will be applied to extensive simulated datasets and forthcoming CSST slitless spectroscopic observations for further cosmological and extragalactic analyses.
\end{abstract}


%%%%%%%%%%%%%%%%%%%%%%%%%%%%%%%%%%%%%%%%%%%%%%%%%%

%%%%%%%%%%%%%%%%% BODY OF PAPER %%%%%%%%%%%%%%%%%%

\section{Introduction}\label{sec:1}
Contemporary cosmology has entered an era of precise measurements of cosmological parameters \citep{2011AIPC.1396...28C,Wallisch2019}. This advancement is predominantly facilitated by sophisticated observational technologies and analytical methodologies, encompassing high-precision measurements of the cosmic microwave background \citep{1965ApJ...142..419P, 2020A&A...641A...6P}, the exploitation of Type Ia supernovae distance-luminosity relationships \citep{1998AJ....116.1009R}, and the investigation of dark matter \citep{1987ARA&A..25..425T} and dark energy \citep{2003RvMP...75..559P} properties through phenomena such as baryon acoustic oscillations \citep{1970ApJ...162..815P}, redshift-space distortions \citep{1987MNRAS.227....1K,2023MNRAS.523.6360W}, and weak gravitational lensing \citep{2001PhR...340..291B}.

The progress in understanding the evolution of our universe is fundamentally tied to vast quantities of astronomical observational data, with large-scale imaging and spectroscopic surveys serving as paramount tools. Over the past decades, there are many of such sky surveys providing a vast number of galaxy redshifts, such as the Two-degree Field Galaxy Redshift Survey (2dFGRS) \citep{2001MNRAS.328.1039C}, the Sloan Digital Sky Survey (SDSS) \citep{2000AJ....120.1579Y}, the Hyper Suprime-Cam Subaru Strategic Program (HSC-SSP) \citep{2018PASJ...70S...4A}, the Dark Energy Survey (DES) \citep{2016MNRAS.460.1270D}, and the Javalambre-Physics of the Accelerated Universe Astrophysical Survey (J-PAS) \citep{2014arXiv1403.5237B}. They have advanced our understanding of the expanding universe and the nature of dark energy. 

Presently, a suite of cosmological surveys has been initiated as the Stage-IV dark energy experiments \citep{2006astro.ph..9591A}, aiming to explore the expansion history and structure growth of the universe. Among these are prominent ground- and space-based surveys, including Prime Focus Spectrograph \citep[PFS;][]{2016SPIE.9908E..1MT}, Dark Energy Spectroscopic Instrument \citep[DESI;][]{2016arXiv161100036D, 2023ApJ...943...68L}, Large Synoptic Survey Telescope \citep[LSST;][]{2009arXiv0912.0201L}, China Space Station Telescope \citep[CSST;][]{2011SSPMA..41.1441Z,2021ChSBu..66.1290L}, Euclid \citep{2011arXiv1110.3193L}, and Roman Space Telescope \citep[RST;][]{2012arXiv1208.4012G}. These projects signify the dawn of a new era in cosmology, characterised by unprecedented precision of cosmological parameter measurements.

The Chinese Space Station Optical Survey (CSS-OS), utilizing the CSST, will launch a new large-scale cosmological survey featuring both multi-wavelength imaging and slitless spectroscopy \citep{2018MNRAS.480.2178C, 2019ApJ...883..203G, 2021ChSBu..66.1290L}. This survey will cover a sky area of approximately 17,500 square degrees with seven imaging bands and three slitless spectroscopic bands. The telescope is forecasted to significantly advance our understanding of the accelerated expansion of the universe. Comparative assessments indicate that CSS-OS, relative to contemporary weak gravitational lensing and galaxy clustering surveys, will significantly enhance the constraints of cosmological parameters \citep{2019ApJ...883..203G}.

The CSS-OS is designed to facilitate high-precision determinations of cosmological parameters by employing both photometric and spectroscopic redshift measurements. This study narrows its focus to redshift derivations extracted from the CSST slitless spectroscopic data, specifically leveraging the diagnostic power of emission lines. The spectroscopic redshift as well as emission line information of galaxies is also essential for studying the galaxy formation and evolution. We specially develop the process for the redshift determination and evaluate the redshift measurement quality using simulated slitless spectra. These spectra are generated by the CSST Emulator for Slitless Spectroscopy \citep[CESS, Paper II;][]{2024MNRAS.528.2770W} using the reference mock galaxy redshift catalogs tying to the photometric data from the DESI Legacy Imaging Surveys (DESI LS) \citep[Paper I;][]{2024MNRAS.529.4015G}.

In this third paper (Paper III) of the series detailing the CSST large scale structure analysis pipeline, the structure of our paper is organized as follows. Section ~\ref{sec:2} introduces the characteristics of the CSST and CSS-OS slitless spectroscopy. The simulation of the slitless spectra is also summarised in this section. Section ~\ref{sec:3} depicts the detailed process of the spectroscopic redshift measurement based on nebular emission lines. In Section~\ref{sec:4}, we apply our redshift measurement method to the simulated CSST slitless spectra to calculate redshifts and assess the redshift quality. The redshift degeneracy due to the emission-line matching 
confusion and success rate of all redshift outputs are discussed in Section \ref{sec:discussion}. Section~\ref{sec:summary} gives the summary. 

\section{Slitless Spectra of CSST Main Sky Survey}\label{sec:2}

\subsection{Main CSST parameters}
CSST is a space-based telescope planed by the China Manned Space Program as outlined in \citet{2011SSPMA..41.1441Z, 2021ChSBu..66.1290L}. It is operated independently yet sharing the same orbit with the China Space Station at approximately 400\,km above Earth, which is convenient for routine maintenance. The telescope has a primary mirror of 2\,m in diameter and a focal length of 28\,m. The radius encircling 80 per cent energy of the point spread function (PSF) is less than 0\arcsec.15, which is comparable to that of the Hubble Space Telescope (HST). CSST is equipped with five scientific instruments, including a survey camera, a terahertz receiver, a multichannel imager, an integral field spectrograph, and a cool planet imaging coronagraph. Among these, the survey camera stands out as the cornerstone of CSST observational capabilities, handling both imaging and slitless spectroscopic observations. It provides a field of view of about 1.1\,deg$^2$. The wavelength coverage spans from $255$ to $1000$\,nm and the average resolution is no less than 200. The image quality of the spectra is specified to be $R_\mathrm{EE80}\le$ 0\arcsec.3. The survey camera is used to conduct the CSS-OS, which utilises 70 per cent of CSST total observation time over a minimum duration of ten years. This ambitious survey aims to map approximately 17,500 square degrees of the sky, focusing on regions with medium-to-high Galactic and Ecliptic latitudes. The CSST together with the onboard survey camera are distinguished by a suite of remarkable characteristics: high spatial resolution (PSF $R_\mathrm{EE80}$ $\leq$ 0\arcsec.15), exclusive capacity for near-ultraviolet spectral observations, wide wavelength coverage ($255-1000$ nm), and large number of filters (18 filters for photometry and 12 filters for slitless spectroscopy) \citep{2021ChSBu..66.1290L,2019ApJ...883..203G}. This powerful combination is poised to significantly advance our understanding of dark energy, facilitate rigorous tests of gravitational theories, and provide robust validations for cosmological models. 

%
%survey is designed to simultaneously capture data in 7 imaging bands and $3$ slitless spectroscopic bands. Impressively, the CSS-OS is set to occupy 70 per cent of CSST's total observation time, spanning a sky area of approximately 17,500 square degrees. CSST boasts several notable advantages, including its exceptional image quality, unique near-ultraviolet band observation capabilities \citep{2021ChSBu..66.1290L}, and enhancements in various aspects such as wavelength coverage, number of filters, and spatial resolution \citep{2019ApJ...883..203G}.

%Scheduled for a decade of scientific exploration, CSST will concentrate primarily on large-scale multi-band imaging and slitless spectroscopic surveys of the sky.

%The CSST is equipped with multiple instruments, among which the main-survey camera will conduct a large-scale optical survey

\begin{table}
	\centering
	\caption{Key parameters of CSST and related slitless spectroscopy}
	\label{Key parameters of CSST}
	\begin{tabular}{cc} % four columns, alignment for each
		\hline
		Parameter & Value \\
		\hline
        Primary Mirror & $2$ m \\
        Focal Length & $28$ m \\
        Imaging quality & PSF $R_\mathrm{EE80}$ $\leq$ 0\arcsec.15 \\
        Field of View & $\geq 1.1$ deg$^2$.\\
        Wavelength coverage & $0.255-1.0\ \mu$m\\
        Exposure Time & 150 s $\times$ 4 \\[0.2cm]
        Spectral Bands & \makecell{GU: 255--400\ nm \\ GV: 400--620 nm\\ GI: 620--1000 nm}\\[0.5cm]
        Spectral Resolution & $\geq 200$ \\[0.2cm]
        \makecell{Limiting magnitude (mag) \\ ($5\sigma$ for point source in AB mag)} & \makecell{GU: $23.2$ \\ GV: $23.4$\\ GI: $23.2$}\\[0.5cm]
        Sky coverage of CSS-OS & 17500 deg$^2$\\
		\hline
	\end{tabular}
\end{table}

%The survey camera consists of the main focal plane and celestial calibration components, filters, slitless spectral components, shutters, and some other elements.
The survey camera at the focal plane contains 30 9k$\times$9k CCD detectors, collectively covering an area of about $1.1^{\circ} \times 1.2^{\circ}$ at the centre of the field of view. Eighteen of these detectors are dedicated to multi-band imaging, covered with filters for NUV (four), $u$ (two), $g$ (two), $r$ (two), $i$ (two), $z$ (two), and $y$ (four) bands. The other twelve detectors are assigned to slitless spectroscopic tasks, employing GU (four), GV (four), and GI (four) gratings. To avoid huge filter wheels in space and enhance efficiency, the filters and gratings are mounted directly in front of each detector without any moving parts. The standard exposure time of 150\,s $\times$ 4 achieves 5$\sigma$ limiting magnitudes for point sources of $23.2$, $23.4$, and $23.2$\,mag in the GU, GV, and GI bands, respectively. Additionally, there are plans to survey multiple deep fields of about 400 deg$^2$ with 6.7 times longer exposures (1.2\,mag deeper). Nonetheless, the present work confines its scope to the wide survey. Table~\ref{Key parameters of CSST} lists some key parameters of the telescope and slitless spectroscopy.

\subsection{Simulation of CSST slitless spectra} \label{sec:simulation}
To evaluate the capacity of the CSST slitless spectroscopic survey for cosmological constraints, we undertake the development of a comprehensive CSST large scale structure analysis pipeline. This pipeline encompasses an entire sequence, including cosmological simulations, spectral emulation, large-scale structure analyses, and cosmological forecasts. Paper I meticulously outlines the methodology behind assembling the reference mock galaxy redshift catalogs from numerical simulations using presupposed cosmological parameters \citep{2024MNRAS.529.4015G}. Following this, Paper II introduces the CESS emulator \citep{2024MNRAS.528.2770W}, which is a pivotal tool designed for the synthesis of slitless spectra from these mock catalogs. Below, we encapsulate the CESS simulation process, and yet encourage readers to refer to \citet{2024MNRAS.529.4015G} and \citet{2024MNRAS.528.2770W} for details.

The CESS takes a high-resolution galaxy model spectrum as input and  produce a 1D slittless spectrum according to the instrument parameters specific to CSST grisms and detectors, such as wavelength range, total throughput, exposure time of 4 $\times$ 150\,s, and pixel scale. In terms of grism resolutions, R = 241, 263, and 270 were set at the centre wavelengths of the three bands GU, GV, and GI, respectively, which met the expected quality. Additionally, the emulator takes into account the galaxy morphology quantified by four parameters, including S\'ersic index ($n$), effective radius ($R_e$), position angle (PA), and axis ratio ($b/a$). These morphological parameters can significantly influence the characteristics of the resulting CSST spectrum. Furthermore, the emulator incorporates various forms of noise, including sky background noise, wavelength calibration error, dark current, and read-out noise. For the sky background estimation, the CESS adopts the data from the HST WFC3 Instrument Handbook \citep{2023wfci.book...15D}, providing a specific value of sky background magnitude for each input source according to the sky coordinates.

% \textcolor{purple}{Specifically, the galaxy morphology will produce self-broadening effects on the spectra, increasing the width of emission lines and reducing their peak intensity. Among the four morphological parameters, $R_e$ has the greatest contribution on the self-broadening effects, followed by $b/a$ and PA, and S\'ersic index $n$ has little impact \citep{2024MNRAS.528.2770W}.}
%Table~\ref{Important Parameters of CESS} outlines some crucial parameters utilised in CESS to reflect the characteristics of the CSST.
%
%\begin{table}
%	\centering
%	\caption{Important Parameters of CESS}
%	\label{Important Parameters of CESS}
%	\begin{tabular}{cccc}

%	   \hline
%       \multirow{2}{3.3cm}{Parameter} & %\multicolumn{3}{c}{Value}\\\cline{2-4}
%       & GU & GV & GI \\\hline
%       Wavelength Coverage (nm)  & 255 – 420 & 400 – 650 & 620 – 1000\\
%       Spectral Resolution & 241 & 263 & 270\\
%       5$\sigma$-depth magnitude (mag)   & 23.2 & 23.4 & 23.2\\\hline
%	\end{tabular}
%\end{table}

The emulator utilizes mock galaxy redshift surveys (MGRSs) sourced from the Jiutian numerical simulations to generate its input model spectra. These MGRSs utilise the DESI LS dataset \citep{2010AJ....140.1868W,2017PASP..129f4101Z,2019AJ....157..168D}, assigning $z$-band magnitudes and redshift values up to $z\sim1.0$ and imposing a magnitude cut of $m_z<21$ mag for galaxies. This magnitude cut is strategically implemented to enhance the accuracy of photometric redshift estimates; specifically, it ensures high-quality multi-wavelength photometry, achieving a photo-z accuracy of approximately $\sigma_{z} / (1+z) \sim 0.02$. Moreover, this cut align roughly with the GI depth of CSST slitless spectroscopy. Given that the GI band is significantly wider than the $z$ band, it allows for a deeper limit of about one magnitude. The depth for extended sources would be over 0.5 mag shallower than that for point sources. While a relatively brighter magnitude cut may reduce the number of high-redshift galaxies, there remains a sufficient number of objects to effectively assess the redshift measurements.
% \textcolor{purple}{The z-band is the reddest one of the DESI LS optical bands, and the reddest band GI of CSST is essential for redshift measurement at $z<1$ \citep{2024MNRAS.529.4015G}, so the magnitude limit cut of $m_z<21$, which is brighter than those of CSST and DESI LS, is chosen to ensure completeness. If CSST will observe darker galaxies, the mock galaxies can be extended by HSC \citep{2018PASJ...70S...4A} and PFS \citep{2014PASJ...66R...1T,2022arXiv220614908G}.}
Following this, a photometric analysis of the spectral energy distribution (SED) is performed using BayeSED \citep{2014ApJS..215....2H,2019ApJS..240....3H,2023ApJS..269...39H} using the five-band DESI photometric data ($g$, $r$, $z$, $W1$, and $W2$).

BayeSED utilizes the simple stellar population models from \citet{2003MNRAS.344.1000B}, the initial mass function of \citet{2003PASP..115..763C}, an exponentially declining star formation history, the dust extinction law of \citet{2000ApJ...533..682C}, and the Cloudy nebular emission model \citep{2017RMxAA..53..385F}. In conducting spectral energy distribution (SED) analysis, the parameters considered include galaxy ages ranging from 10 Myr to 13.46 Gyr, star formation timescales $\tau$ from 1 Myr to 1000 Gyr, stellar metallicities from 0.005 $Z_{\odot}$ to 5 $Z_{\odot}$, and extinction coefficients $A_v$ from 0 to 4. These parameters are sampled within their respective ranges using the Multinest sampling algorithm, employing a Bayesian approach. Additionally, BayeSED incorporates the nebular continuum and emission lines into the simple stellar population model for stellar populations younger than 10 Myr, based on the geometry and parameters outlined in the photoionization model of nebulae by \citet{2017ApJ...840...44B}. This allows for the calculation of composite stellar population synthesis models, yielding energy spectra that account for both stellar populations and gas. For the young stellar populations under 10 Myr generating emission lines, several assumptions are made: the nebular ionisation parameter log($U$) = $-2.3$, inner radius log($r_{in}$/cm) = 19, hydrogen number density log($n_{\rm{H}}/\rm{cm^{-3}}$) = 2, and gas metallicity log($Z_{\rm{gas}}/Z_{\rm{star}}$) = 0, which is the same as the metallicity of the stellar population. This methodology results in a comprehensive library of over $10^8$ galaxies, each accompanied by a best-fit high-resolution model spectrum and other relevant physical properties. For more detailed information on spectral simulation, we refer readers to \citet{2024MNRAS.528.2770W} and the references cited within.
% \textcolor{purple}{These model spectra are the intrinsic values of the simulated slitless spectra, but due to the introduction of random noise during the simulation, there may be some discrepancies of fluxes between the simulated spectra and model spectra.}

\begin{figure}
	\includegraphics[width=\columnwidth]{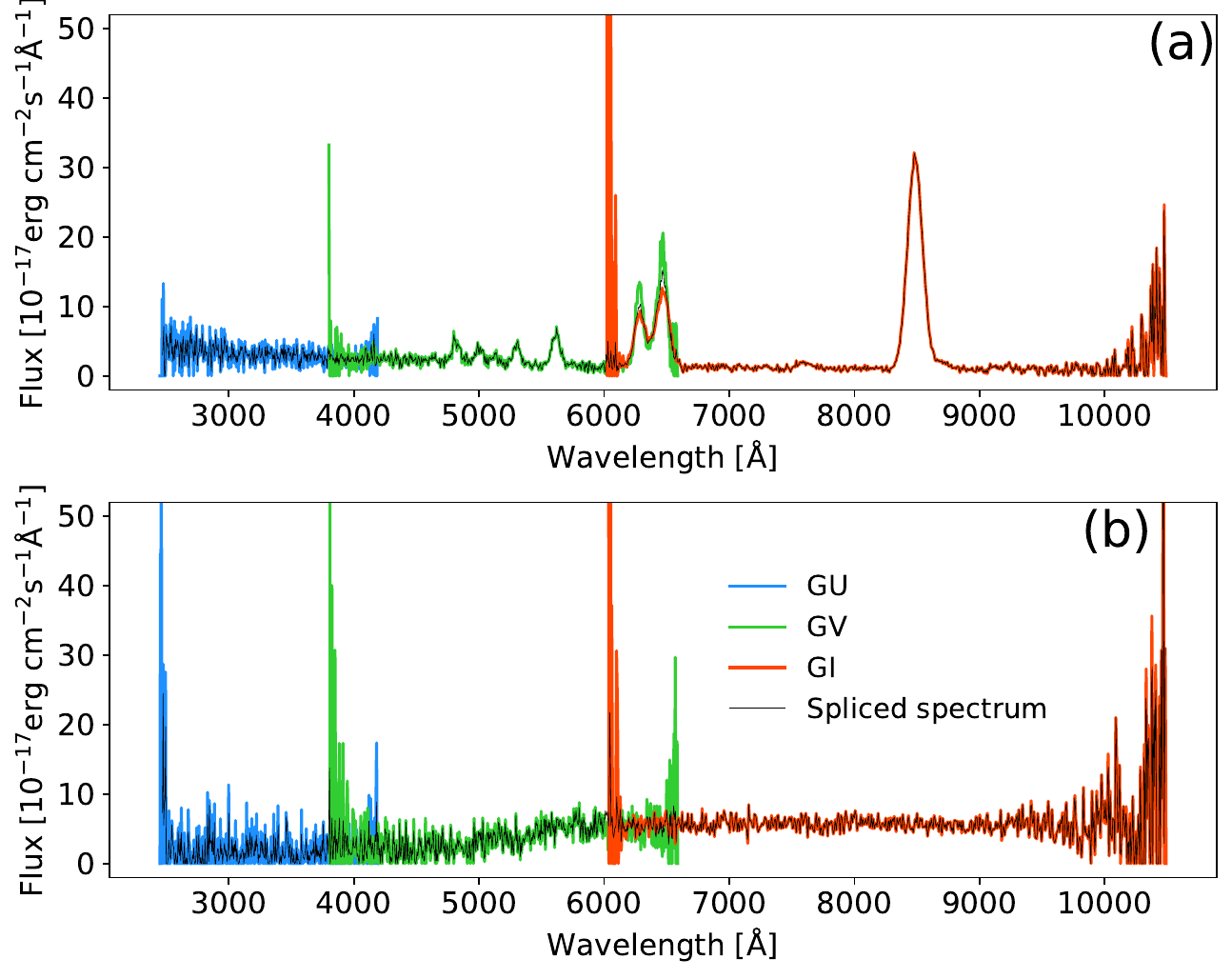}
    \caption{Examples of simulated CSST slitless spectra. The solid blue, green, and orange lines represent the GU, GV, and GI spectra, respectively.
    % The black dashed lines are the theoretical flux errors within these three bands.
    The black solid lines represents the spliced spectra. The spectrum in the upper panel exhibits clear emission line characteristics, whereas the one in the lower panel does not display prominent emission lines.}
    \label{original_spectra}
\end{figure}

Fig.~\ref{original_spectra} displays two example spectra from the simulated dataset, each comprising three distinct bands. The spectrum depicted in the upper panel exhibits prominent emission lines, whereas the one in the lower panel lacks clear, identifiable emission lines. The significant fluctuations at the two ends of each spectroscopic channel result from the low instrumental response. For the following analyses of this paper, we combined the spectra from the three channels into a spliced spectrum in an error-weighted manner. 
%\textcolor{purple}{, and adjusted the interval between adjacent pixels to 5 \AA}

\begin{figure}
	\includegraphics[width=\columnwidth]{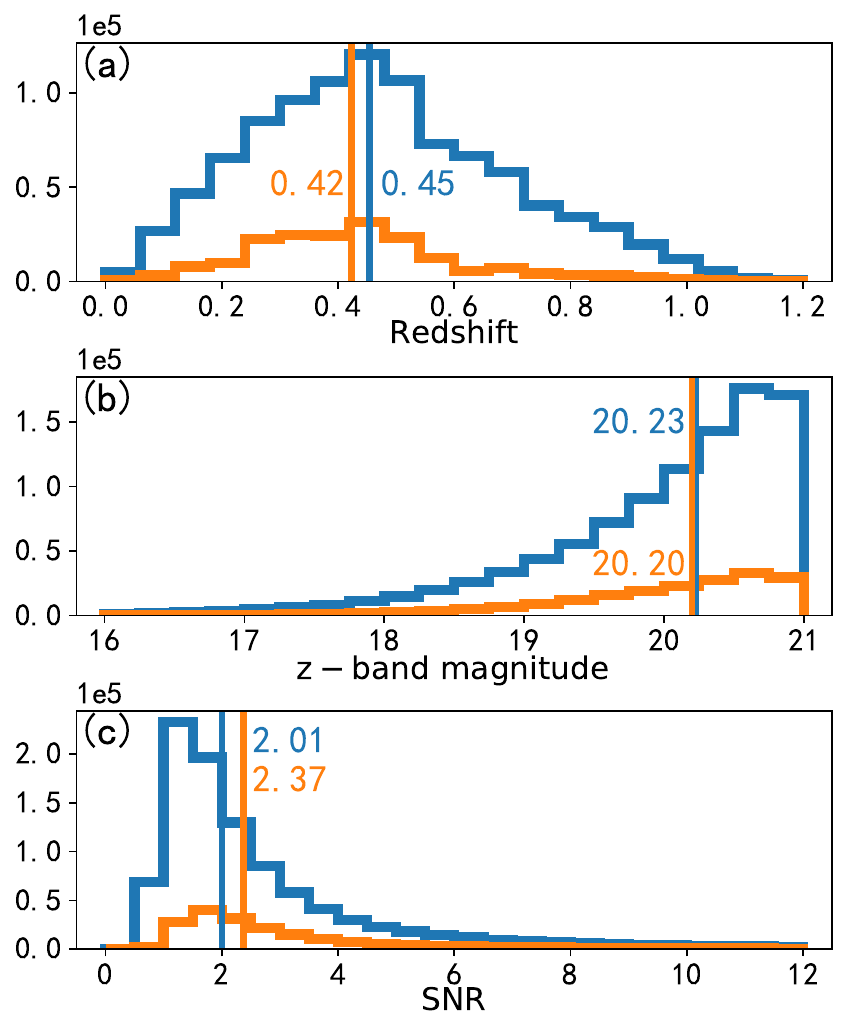}
    \caption{Distributions of parameters for all CESS simulated spectra (blue) for redshift (a), $z$-band magnitude (b), and SNR per pixel (c). The orange histograms indicate the parameter distributions for spectra with successful redshift measurements (see Section \ref{sec:redshift_results}). Median values are indicated by vertical lines.} \label{data_properties}
\end{figure}

For our redshift analysis, a random subset of approximately $10^6$ spectra is generated by the CESS emulator. Fig.~\ref{data_properties} provides an overview of key parameter distributions: redshift, $z$-band magnitudes, and median spectral signal-to-noise ratio (SNR) per pixel. The redshift distribution peaks at a median value of $0.45$. The $z$-band magnitudes range from $17$ to $21$\,mag, with a median value of about $20.23$\,mag. Concerning the spectral quality, the SNR distribution skews towards lower values with a median SNR of $2$, showing the typical noise levels in our spectral dataset.

\section{Process of emission-line redshift measurement}\label{sec:3}

\subsection{Emission lines}
\begin{figure}
	\includegraphics[width=\columnwidth]{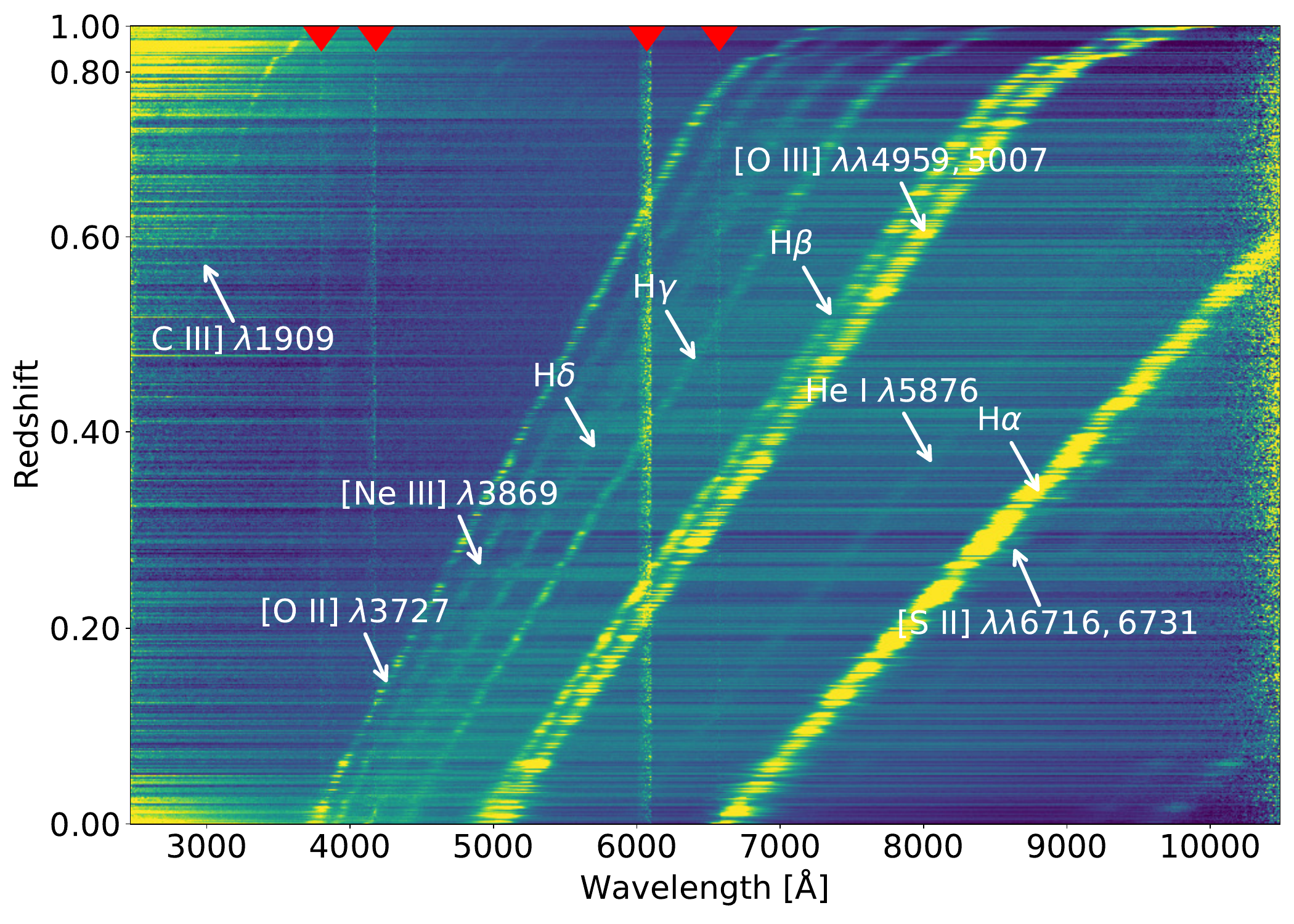}
    \caption{Stacked slitless spectra of ELGs, showing the redshifts spanning from 0 to 1, arranged in ascending order from bottom to top. Prominent emission lines are labelled. The flux of each spectrum has been normalised to highlight these emission lines. Notably, the slitless spectra of CSST encompass three separate bands, leading to observable seams between adjacent bands. The presence of low SNRs at the ends of the spectral bands primarily causes these seams.  The four red arrows at the top point out the positions of these seams for reference.}
    \label{EL_position_vs_redshift}
\end{figure}

Emission line galaxies (ELGs), characterised by their distinctive spectral emission features originating from ongoing star formation activities, facilitate precise redshift determinations by comparing the observed line wavelengths with their intrinsic counterparts. Fig.~\ref{EL_position_vs_redshift} showcases the stacked spectra of $1,629$ ELGs selected from the CESS simulated data. It illustrates their spectral behavior across a redshift interval of $0<z<1$. The visualization demonstrates a shift towards longer wavelengths for the most prominent emission lines as redshift increases. Markings within the figure highlight these emission lines, displaying varying intensities in line with established ratios. These characteristic emission patterns serve as crucial elements for precise redshift determinations.

Table~\ref{emission_line_templates} lists the key emission lines, serving as all optional template lines for redshift determination within the spectral coverage of the CSST slitless grism. The table outlines the detectable redshift ranges and corresponding wavelengths\footnote{\url{http://astronomy.nmsu.edu/drewski/tableofemissionlines.html}} of the template emissions. Notably, for lines such as $\text{C IV} \lambda\lambda1548,1551$, $\text{Mg II]} \lambda\lambda2796,2803$, and $\text{[S II]} \lambda\lambda6716,6731$, the given wavelengths represent the averages of the two lines. Additionally, the wavelength of $\text{[O III] } \lambda\lambda4959,5007$ is derived as a weighted average based on the flux ratio of 1:2.98 \citep{2023zndo...8210732B}. In light of challenges posed by noisy data in the GU band, as highlighted in Fig.~\ref{EL_position_vs_redshift}, certain lines like Ly$\alpha$ and C IV$\lambda\lambda$1548,1551 may be difficult to distinguish effectively. Furthermore, lines such as [Ne III]$\lambda$3869, H$\delta$, and He I$\lambda$5876 are relatively faint. As a result, this study primarily focuses on prominent lines, including $\text{[O II] } \lambda3727$, H$\gamma$, H$\beta$, $\text{[O III] } \lambda\lambda4959,5007$, and H$\alpha$. We also consider the weaker $\text{C III]}\lambda1909$, since it is the sole relatively prominent ultraviolet emission line as shown in Fig.~\ref{EL_position_vs_redshift}. These lines are specifically marked in Table~\ref{emission_line_templates}. The additional lines in this table could be beneficial for deeper slitless spectroscopic observations.

\begin{table}
	\centering
	\caption{Key emission lines that can be used for redshift determinations}
	\label{emission_line_templates}
	\begin{tabular}{ccc}
		\hline
		Line name & Wavelength (\AA) &  Redshift coverage\\
		\hline
        $\text{Ly} \alpha$ & $1215.670$ & $1.10-7.23$ \\
        $\text{C IV} \lambda\lambda1548,1551$ & $1549.480$ & $0.65-5.45$ \\
        $\text{C III]} \lambda1909^{\star}$ & $1908.734$ & $0.34-4.24$ \\
        $\text{Mg II]} \lambda\lambda2796,2803$ & $2799.117$ & $0-2.57$ \\
        $\text{[O II]} \lambda3727^{\star}$ & $3727.424$ & $0-1.68$ \\
        $\text{[Ne III]} \lambda3869$ & 3868.760 & $0-1.58$\\
        %$\text{[Ne III]} \lambda3967$ & 3967.470 & $0-1.52$\\
        $\text{H}\delta$ & 4101.742 & $0-1.44$\\
        $\text{H}\gamma^{\star}$ & $4340.471$ & $0-1.30$\\
        $\text{H}\beta^{\star}$ & $4861.333$ & $0-1.06$\\
        $\text{[O III]} \lambda\lambda4959,5007^{\star}$ & $4994.800$ & $0-1.00$ \\
        $\text{He I} \lambda5876$ & 5875.624 & $0-0.70$\\
        $\text{H}\alpha^{\star}$ & $6562.819$ & $0-0.52$ \\
        $\text{[S II]} \lambda\lambda6716,6731$ & $6723.625$ & $0-0.49$ \\
		\hline
	\end{tabular}
	    \begin{tablenotes}
            \footnotesize
            \item[1] $^{\star}$These emission lines are used for redshift determination in this work.
        \end{tablenotes}
\end{table}

\subsection{Basic flowchart}
\begin{figure*}
	\includegraphics[scale=0.165]{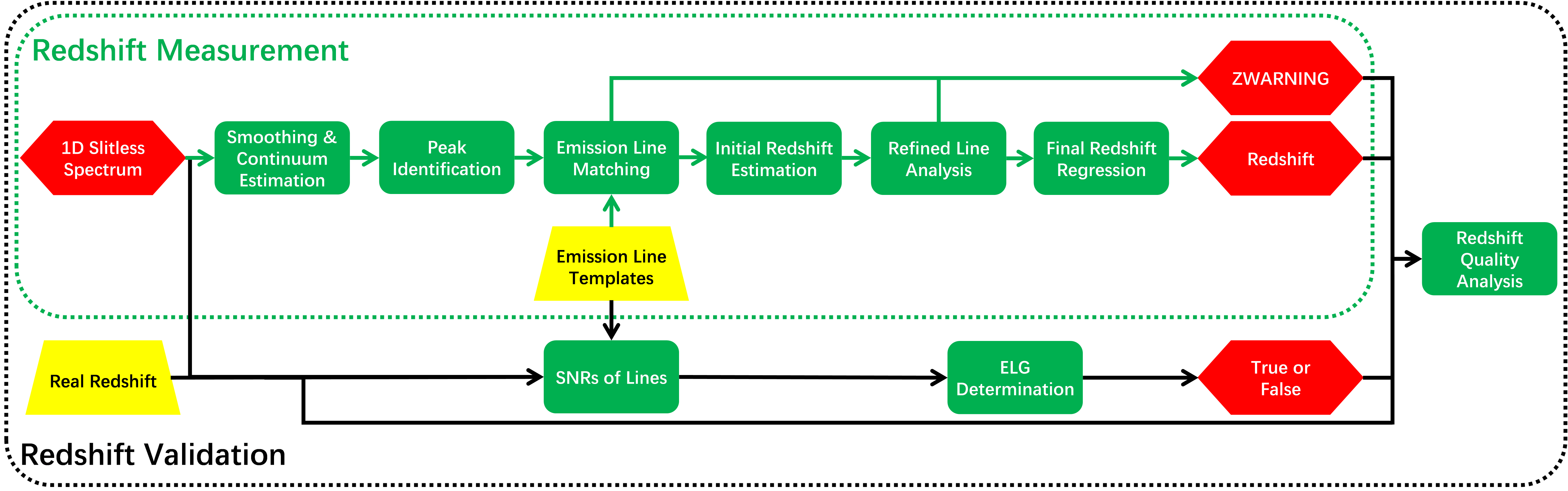}
    \caption{Schematic flowchart of the redshift measurement based on emission lines for CSST slitless spectra.}
    \label{flow_chart}
\end{figure*}

Fig.~\ref{flow_chart} illustrates the tailored redshift measurement methodology designed for ELGs. It presents the detailed workflow. The sequence encapsulated within the green dashed boundary delineates the core redshift estimation process. The process starts with a 1D slitless spectrum containing wavelengths, fluxes, and corresponding uncertainties. The initial stages involve spectral smoothing and continuum evaluation. Subsequently, peak identification isolates potential emission lines. By comparing these peaks with a reference set of rest-frame emission-line wavelengths, initial redshift solutions are obtained. A minimum of two identifiable emission lines is necessary for the redshift calculation. To refine these estimates, Gaussian fitting is utilised to enhance the accuracy of central wavelength determinations. Ultimately, three top-ranked redshifts are yielded through proper ranking criteria. Each redshift estimation is accompanied by a ZWARNING flag (see Section \ref{sec:zwarning}) highlighting any potential measurement issues. For validation purposes, the entire process outlined in both the green and black boxes of Fig.~\ref{flow_chart} is executed. This validation involves calculating SNRs at anticipated emission line wavelengths and assessing redshift quality using the mock galaxy dataset. Here, ELGs are identified based on the presence of at least two emission lines with fluxes surpassing a minimum SNR threshold of 3. 

The process for redshift determination is delineated as follows:
\begin{enumerate}
  \item[(1)] \textit{Smoothing and Continuum Estimation}: Commence by obtaining the wavelength, flux, and associated error information from a one-dimensional slitless spectrum. Utilise Gaussian filtering for flux smoothing and estimate the continuum through a combination of median and Gaussian filtering methods.
  \item[(2)] \textit{Peak Identification}: Adopt a peak detection algorithm (\textit{find\_peaks}\footnote{\url{https://docs.scipy.org/doc/scipy/reference/generated/scipy.signal.find_peaks.html}}) to identify potential emission lines within the smoothed flux spectrum. These lines are identified as those peaks with smoothed fluxes sufficient above the surrounding continuum.  
  \item[(3)] \textit{Emission Line Matching}: Match the identified candidate lines to theoretical emission line templates, and propose potential redshifts. Discard single line matches and remove duplicated matching results. This process can produce several combinations of emission lines that correspond with template lines, resulting in multiple potential redshift outputs.
  \item[(4)] \textit{Initial Redshift Estimation}: For surviving matches, calculate a preliminary redshift through linear regression of the observed peaks versus template emission line wavelengths. 
  \item[(5)] \textit{Refined Line Analysis}: Enhance precision by applying Gaussian profiles to fit each emission line, extracting more accurate line centres and additional parameters (e.g. flux, width, and equivalent width). Multi-Gaussian profiles are used for lines that are in close proximity to one another.
  \item[(6)] \textit{Final Redshift Regression}: Re-evaluate the central wavelengths after Gaussian fitting and undertake another linear regression for refined redshift calculation. During this process, outliers from matched line sets are removed using sigma clipping.
  \item[(7)] \textit{Redshift Ranking and Output}: Organize all candidate redshifts based on the precision of Gaussian fits and output the top three, if available, alongside their respective errors, number of matched lines, and quality flags of ZWARNING. (e.g., whether the flux ratio of specific emission lines exceeds the predetermined range).
  
  %\item[(8)] Sort all redshifts based on the Gaussian fitting of emission line wavelengths and output the top three selected redshifts (if available), redshift errors, ZWARNING flags, the number of matched lines, and the number of lines used in redshift fitting. The first listed redshift represents the optimal one.
\end{enumerate}

The basic inputs include:
\begin{itemize}
  \item \textit{wave}: Array of wavelengths constituting the input spectrum.
  \item \textit{flux}: Corresponding flux values in the spectrum.
  \item \textit{error}: Measurement uncertainty associated with each flux value.
  \item templates: A list specifying the employed template emission lines for matching, with all feasible options detailed in Table~\ref{emission_line_templates}. 
  \item \textit{emission\_line\_threshold}: A defined threshold to discriminate significant emission lines during detection.
  \item \textit{redshift\_upperlimit}: An optional ceiling for considered redshift values; any redshifts surpassing this limit are disregarded. 
  \item \textit{real\_redshift}: The actual redshift of the spectrum, which is for testing purposes only.
\end{itemize}

Output deliverables include:
\begin{itemize}
    \item The top three calculated redshifts.
    \item Associated errors for these redshift estimates.
    \item ZWARNING flags for these redshift measurements, indicating any potential issues.
    %\item The count of emission lines matched for each of the top three redshifts.
    \item Number of matched emission lines for determining each of the top three redshifts.
    \item $\chi^2$ of the redshift fitting.
    \item Emission line properties, including central wavelengths, fluxes, flux errors, line width, equivalent width, etc.
\end{itemize}

\subsection{Emission-line identification and template matching} \label{sec:em_match}
To detect emission lines in the CSST slitless spectra, we begin by smoothing the spectra and estimating the underlying continua. The smoothing process aims for suppressing the noise while retaining essential spectral features. We achieve this by applying a Gaussian filter using a kernel with the standard deviation of 5 pixels. Note that the wavelength element is about 5 \AA\ per pixel and the resolution is about 20 \AA\ in GV band. The smoothed spectra are used for identifying the emission lines. Our approach to estimating the continua involves a two-step filtering process: a median filter with a window size of 251 pixels and a Gaussian filter with a kernel with the standard deviation of 10 pixels. The latter additional smoothing pass helps to create a more refined estimate of the continuum.

In our analysis, we employ the \textit{find\_peaks} Python module to identify potential emission lines in the smoothed spectrum. This process involves detecting peaks within the spectrum, which are indicative of emission features. To enhance the reliability of our peak detection, we configure several parameters within the \textit{find\_peaks} module. Firstly, a minimum separation of 35 pixels between neighboring peaks is established based on the wavelength distance of template emission lines in the CSST slitless spectra. This separation criterion helps in distinguishing individual peaks and avoiding overlap between nearby fake features. Additionally, we set a minimum width of peaks to 9 pixels which is adapted to the line spread functions of the CSST grisms. This parameter aids in filtering out noise spikes that do not represent genuine emission features and is specifically tailored to the spectral resolution. Furthermore, the minimal prominence in the \textit{find\_peaks} module is defined as $0.05$ times the median flux, which means that the differences between the peak and the baseline height on both sides cannot be less than this value. 
In addition, the minimal height of peaks is dynamically determined by calculating the sum of the continuum level and a threshold multiplied by the flux error (\textit{continuum} + \textit{threshold} $\times$ \textit{flux\_error}). This adaptive approach ensures that peaks are detected reliably across varying SNR. Specifically, when the spectral SNR is below 3, the threshold value is set to 0.7. This low threshold is configured to ensure the detection of emission lines with low SNR.  As the SNR increases, the threshold value scales logarithmically. By adjusting these parameters based on the spectral characteristics and noise levels, we aim to enhance the precision and robustness of our peak detection methodology for spectra with varying qualities. For galaxies, particularly quiescent ones at bright magnitudes and high signal-to-noise ratios (SNRs), mis-estimated continua due to absorption features may cause emission line misidentification.
% fluctuations in the continua may be misidentified as emission lines.  
Distinguishing between genuine emission lines and these absorption features on the continuum is challenging with the current peak-finding procedure.  However, this issue could be significantly improved by modeling the spectra using stellar population synthesis models, which will be explored in future work (Wei et al. in preparation).

\begin{figure}
	\includegraphics[width=\columnwidth]{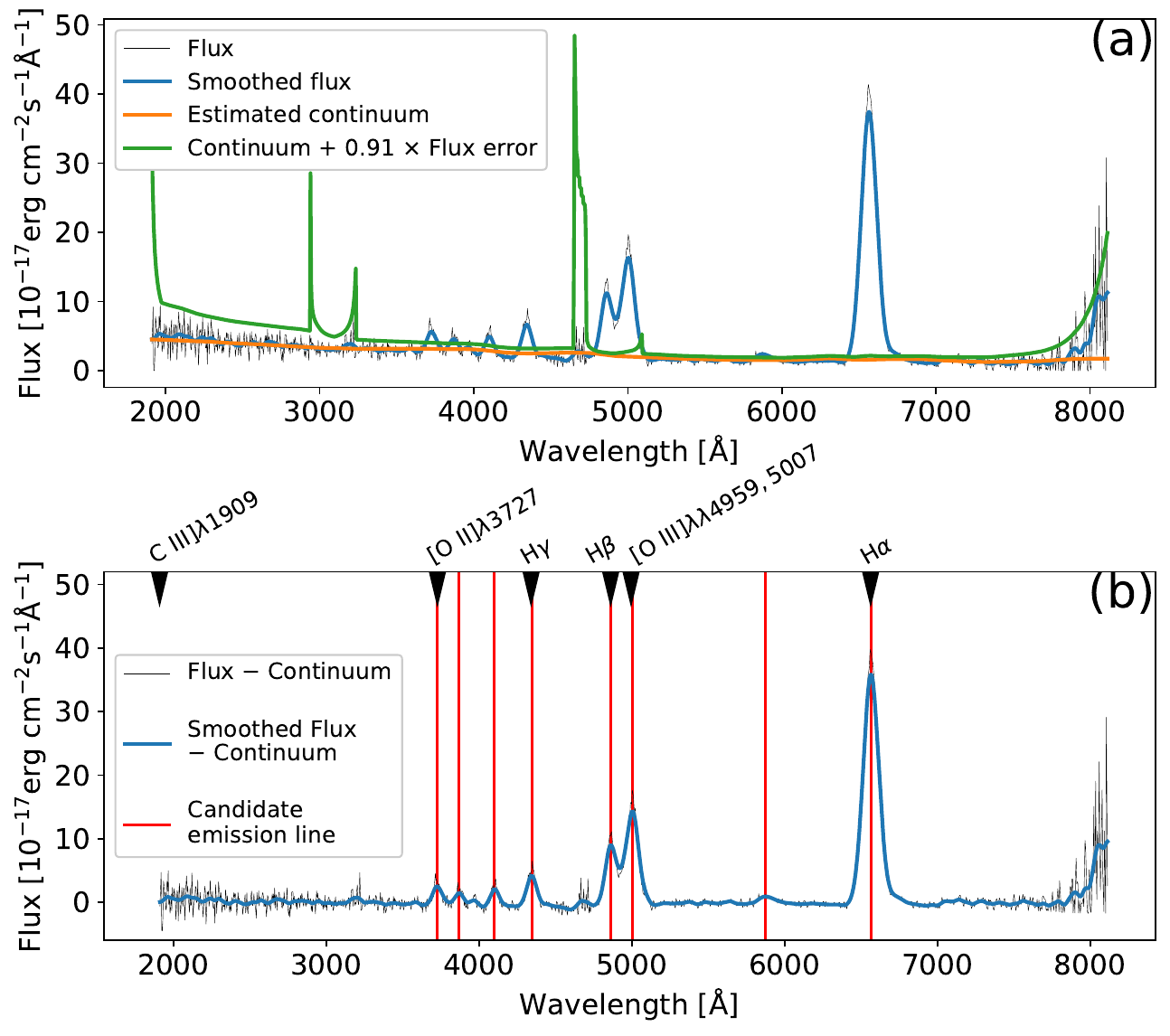}
    \caption{A successful line detection and match. In panel (a), the black curve represents the rest-frame observed spectrum. The blue curve denotes the smoothed spectrum, while the orange curve illustrates the estimated continuum. The green curve indicates the minimum required height for identifying emission lines. Panel (b) displays the emission line identification results. Note that the continuum has been subtracted from the spectrum. The red solid lines highlight the detected candidate emission lines, and black triangles mark the six default template emission lines. }
    \label{a_successful_redshift_measurement_case}

\end{figure}

After identifying candidate emission lines, the subsequent process involves matching these candidates with template lines to determine their rest-frame wavelengths and associated redshift values. To determine redshifts, the identified emission lines are cross-correlated with template emission lines in the following manner: 1) each detected emission line is compared with each template line to establish a potential redshift value; 2) a group of at least two emission lines with approximate redshifts is identified within a redshift tolerance of 0.01; 3) an initial redshift is computed using linear regression between detected line peaks and templates ($\lambda_{\rm{obs}}=(1+z)\lambda_{\rm{rest}}$, where $\lambda_{\rm{obs}}$ and $\lambda_{\rm{rest}}$ represent the observed and template rest-frame wavelengths, respectively and $z$ is the regression coefficient). Through this method, all initial redshift values are derived. The example of a successful line detection and match is illustrated in Fig.~\ref{a_successful_redshift_measurement_case}.

%\boldtext{How the potential redshift are ranked? Please describe detailedly}. 

%We have chosen several prominent spectral emission lines typically found in galaxies as templates for matching and subsequent redshift calculations. These templates cover 14 emission lines, and their redshift ranges are determined based on the wavelength coverage of the CSST slitless spectrum, spanning from $255nm$ to $1000nm$. Table~\ref{emission_line_templates} summarises these templates. The template emission lines encompass prominent lines visible in 

\begin{figure}
	\includegraphics[width=\columnwidth]{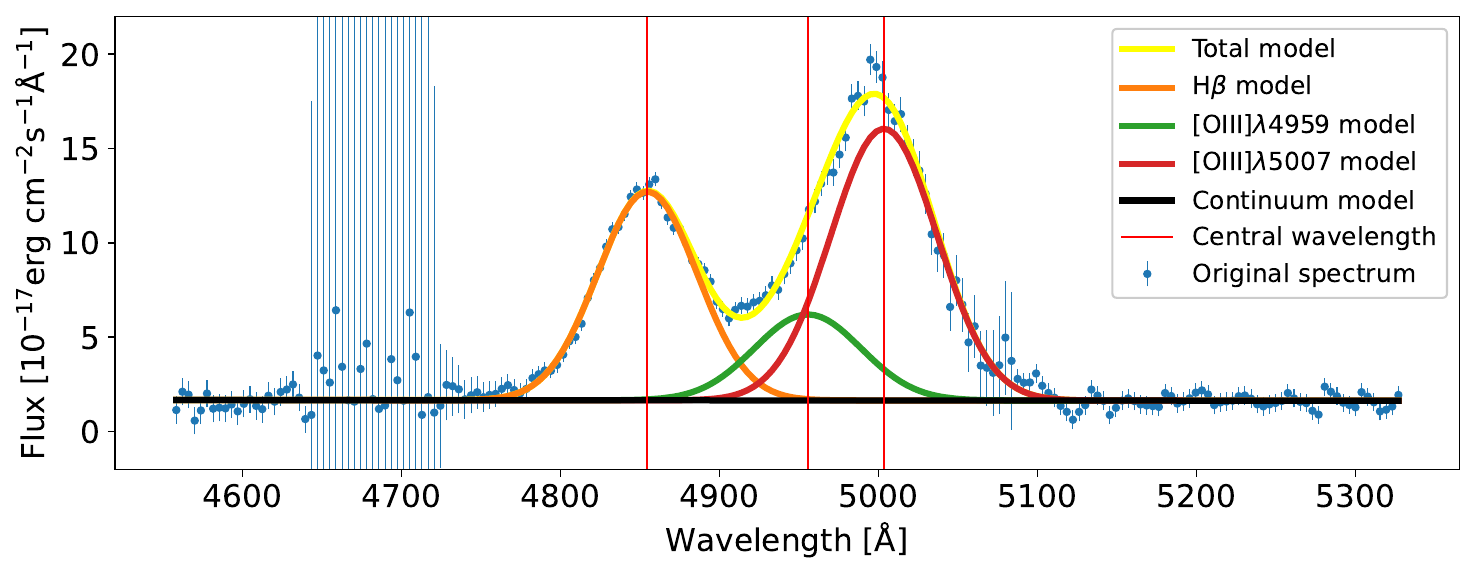}
    \caption{Modeling the emission-line profiles within the wavelength range covered by $\text{H} \beta$, $\text{[O III]} \lambda4959$, and $\text{[O III]} \lambda5007$. The blue points with errors represents the fluxes and associated errors. The orange, green, and red curves are the fitted Gaussian profiles for the $\text{H} \beta$, $\text{[O III]} \lambda4959$, $\text{[O III]} \lambda5007$ emission lines, respectively. The black line is the fitted linear background. The yellow curve present the total fitted model. The three red vertical solid lines mark the central wavelengths of the three emission lines.}
    \label{Gaussian_fitting}
\end{figure}

To enhance the precision of central wavelengths, for every initial redshift, Gaussian fitting is applied to the matched emission lines. We extract a part of the spectrum near each matched emission line, based on the wavelengths of the template emission lines and the initial redshift, and fit both the emission line and the continuum within 200 pixels (5 \AA\ per pixel) in the vicinity of the emission line. When adjacent emission lines are present, multiple Gaussian functions are employed to concurrently fit the profiles of these lines. The background is modelled with a linear function. It is important to emphasize that the final redshift estimation is based solely on the matched template emission lines. This approach is implemented to ensure accuracy in the redshift determination process. The following three cases are handled in this way, including 1) $\text{[O II]} \lambda3727$, $\text{[Ne III]} \lambda3869$, and $\text{[Ne III]} \lambda3967$; 2) $\text{H} \beta$, $\text{[O III]} \lambda4959$, and $\text{[O III]} \lambda5007$; 3) $\text{H} \alpha$ and $\text{[S II]} \lambda\lambda6716,6731$.  It should be noticed that [O II] doublet and [S II] doublet are treated as individual lines due to the limited spectral resolution of the CSST slitless spectroscopy. Additionally, the central wavelengths of $\text{[O III]} \lambda\lambda4959,5007$ are determined through a weighted average considering their respective fluxes. Fig.~\ref{Gaussian_fitting} showcases the profile fitting process for $\text{H} \beta$, $\text{[O III]} \lambda4959$, and $\text{[O III]} \lambda5007$. All remaining matched emission lines are fitted using a single Gaussian profile. 

It is important to mention that we did not account for the morphological broadening effect, as our main objective is redshift determination. In our analysis, the line-spread function is taken into account when establishing the lower bound for the widths of detected peaks. We treat the width as a free parameter during line detection and Gaussian fitting to accommodate both galaxy morphological broadening and possible AGN emissions in real data. We expect to enhance our accuracy with real CSST data in the future by incorporating a spatially varying line-spread function, as well as considering galaxy morphology and AGN classification.

\subsection{Predefined ranges of line ratios to mitigate the degeneracy}
The flux intensities of emission-line galaxies often exhibit strong correlations due to underlying physical conditions of ionization sources \citep{2019ARA&A..57..511K}. Incorrect combinations of emission lines can be effectively eliminated via validating their flux ratios. This verification step plays a crucial role in alleviating degeneracies during the emission line matching process \citep{2023zndo...8210732B}, particularly when the number of matched lines is limited. Actually, well defined emission line ratios are provided in a Python code \textit{Grizli}\footnote{\url{https://grizli.readthedocs.io/en/latest/api/grizli.utils.get_line_wavelengths.html}}. While such intrinsic line flux ratios can vary due to different physical conditions within galaxies, we must also account for the abundance of low-SNR spectra to achieve reasonably accurate redshift measurements. The low SNR often results in significant fluctuations in the observed line ratios. Consequently, we impose quite loose constraints on these line ratios to mitigate redshift degeneracy caused by various line combinations, explicitly marking those redshifts derived from completely nonphysical ratios. When certain emission lines in a spectrum display flux ratios beyond predefined ranges, the corresponding redshift is flagged with a specific ZWARNING designation (see Section \ref{sec:zwarning}). It is worth noting that when the number of matched emission lines is sufficiently large (e.g., $\geq 4$), the redshift determination is typically accurate without the need for flux ratio constraints. In such cases, the flux ratios do not constrain the redshift estimation process. 

% In addition, abnormally small or large line widths might be derived during the Gaussian fitting. The line width can be also utilised to remove those abnormal emission lines.

%We also observed that the number of matched emission lines typically provides a more significant indicator of successful emission line matching compared to flux ratios. It is found that the number of matched emission lines is no less than 4, it strongly suggests a correct matching result. Therefore, to circumvent potential issues stemming from a low signal-to-noise ratio or failed Gaussian fitting, the emission line ratio check is omitted when the number of matched emission lines $\geq 4$.

\subsection{Redshift warning flag} \label{sec:zwarning}
Throughout the redshift measurement process, encountered issues are systematically categorised and documented using the ZWARNING flag. This flag employs a coded system based on powers of 2 ($2^N$), with $N$ representing a positive integer corresponding to a specific type of issue. These issue types are detailed in Table~\ref{ZWARNING_table}. The magnitude of $N$ inherently reflects the severity of the encountered
problem. The aggregation of $\textrm{ZWARNING} = \sum 2^N$ consolidates multiple flags into a solitary numeric summary. By interpreting this sum as a binary representation, each bit directly corresponds to a unique issue, where `1' indicates the presence and `0' signifies the absence of the specific problem.  Generally, the higher the value of $N$, the more frequent the occurrence of redshift measurement failure for this issue.

\begin{table}
	\centering
	\caption{Redshift warning flags}
	\label{ZWARNING_table}
	\begin{tabular}{  m{0.3cm} | m{7.0cm} }
		\hline
        Bit & Description \\\hline
        $0$ &  The Gaussian profile standard deviation of an emission line reaches the preset maximum value of 150 \AA. \\\hline
        $1$ & Observed flux ratio of $\text{H} \alpha$/$\text{[S II]}$ $< 1$. \\\hline
        $2$ & Observed flux ratio of $\text{H} \beta$/$\text{[O III]}$ $> 2$ or  $< 0.02$. \\\hline
        $3$ & Observed flux ratio of $\text{[O II]}$/$\text{[Ne III]}$ $< 0.3$. \\\hline
        $4$ & Observed flux ratio of $\text{H} \alpha$/$\text{H} \beta$ $< 1.5$.  \\\hline
        $5$ & Observed flux ratio of $\text{[O III]}$/$\text{[O II]}$ $< 0.2$. \\\hline
        $6$ & Emission lines excluded due to the failure of Gaussian fitting.  \\\hline
        $7$ & Observed flux ratio of $\text{H}\beta$/$\text{H}\gamma$ $< 1$.  \\\hline
        $8$ & Neither of the two most prominent candidate emission lines could be matched successfully. \\\hline
        $9$ & Observed flux ratio of $\text{H} \alpha$/$\text{H} \gamma$ $< 2$.  \\\hline
        %$10$ & Emission lines excluded due to excessively low SNRs. & $nan$ \\\hline
        %$11$ & Emission lines excluded due to outliers of linear regression. & $nan$ \\\hline
	\end{tabular}
\end{table}

\section{Quality Analysis of Redshift Measurements}\label{sec:4}

\subsection{Quality indicators of redshift measurements}\label{sec:4.1}
The redshift measurement process detailed in Section~\ref{sec:3} is subjected to extensive testing to ascertain the accuracy and reliability of its performance using the simulated CSST slitless spectra described in Section~\ref{sec:2}. Key metrics employed in this evaluation encompass:

\begin{itemize}
  \item \textit{Bias} ($\Delta z_\mathrm{norm}$): measures the systematic offset between the measured and true redshifts. Formulated as $\Delta z_\mathrm{norm} = \frac{z_{Measured}-Redshift}{1+Redshift}$, where $z_{Measured}$ is the code-derived optimal redshift, and $Redshift$ is the input simulation redshift.
%   \item \textcolor{purple}{\sout{Dispersion ($\sigma_{\Delta z_\mathrm{norm}}$): quantifies the scatter in normalized redshift residuals after sigma-clipping to exclude outliers, focusing only on successful measurements ($\left| \Delta z_{norm} \right| < 0.01$).}}
  \item \textit{Normalised median absolute deviation} ($\sigma_\mathrm{NMAD}$): offers a robust measure of scatter, calculated as $\sigma_\mathrm{NMAD} = 1.48 \times median \left| \Delta z_\mathrm{norm} - median(\Delta z_\mathrm{norm}) \right|$, focusing only on successful ones.
%   \item \textcolor{purple}{\sout{\textit{Outlier rates} ($P_{0.01}$): the fraction of measurements deviating significantly from the true redshifts. $P_{0.01}$ measures the percentage of $\left| \Delta z_\mathrm{norm} \right|$ exceeding 0.01.}}
  %where $N_{tot}$ is the number of spectra used for the quality analysis, $N (\left| \Delta z_{norm} \right| > 0.15)$ and $N (\left| \Delta z_{norm} \right| > 3 \sigma_{\Delta z_{norm}})$ are the number of spectra with $\left| \Delta z_{norm} \right| > 0.15$ and $\left| \Delta z_{norm} \right| > 3 \sigma_{\Delta z_{norm}}$, respectively.
  
  \item \textit{Completeness}: reflects the fraction of measurements achieving specified accuracy, specifically where $\left| \Delta z_\mathrm{norm} \right| < 0.01$, out of all spectra analysed. 

  %where represents the number of spectra where the optimal redshift satisfies the condition $\left| \Delta z_{norm} \right| < 0.01$, and $N_{tot}$, while $N_{tot}$ denotes the total count of spectral samples. For the purposes of this paper, a redshift measurement is deemed successful if the optimal redshift meets the criterion $\left| \Delta z_{norm} \right| < 0.01$. Conversely, any redshift measurement failing to meet this standard is considered a failed measurement. Since the code generates three sets of outputs, it should be noted that, unless explicitly stated otherwise, all referenced output results correspond to the optimal values.
  
  \item \textit{Purity}: represents the proportion of successful measurements ($\left| \Delta z_\mathrm{norm} \right| < 0.01$) among all spectra with estimated redshifts.
  
  %\begin{center}
  %    Purity $= N (\left| \Delta z_{norm} \right| < 0.01) / N_{result}$
  %\end{center}
  %where $N_{result}$ denotes the number of spectra from which redshift measurement results could be obtained (it should be noted that for certain spectra, emission line redshift measurement results cannot be determined).
\end{itemize}
Unless otherwise specified, references to output redshifts pertain exclusively to the optimal values generated by our redshift measurement method.

Furthermore, it is important to emphasize that the configuration yielding the redshift measurement performance in this work has been calibrated specifically for simulated spectra. Consequently, when transitioning to analyse actual observational data from the CSST slitless spectroscopy, a recalibration or tuning of the parameters is imperative to ensure continued accuracy and effectiveness. For instance, in our simulations, the resolution of the grisms is set to fixed values, as described in Section~\ref{sec:simulation}. For real spectra, if the resolution differs, we will re-tune the standard deviations of the Gaussian kernels and the window size of the median filter used for smoothing the spectra and estimating the continuum. Additionally, we will adjust the minimum width for emission line identification to ensure accurate detection. Moreover, real spectra often exhibit different noise characteristics compared to the idealized noise in the simulations. Therefore, we will adjust the threshold for emission line identification and the minimum prominence required for detection based on the observed noise levels. These re-tuning and recalibration can be done using a complete sample of galaxies with accurate spectroscopic redshifts, which will help fine-tune the parameters for real data. As mentioned in \citet{2024MNRAS.528.2770W}, a constant spectral resolution is adopted for each of the CSST gratings in the simulations. The corresponding spectral resolutions are approximately 14 \AA, 20\AA, and 30 \AA\ for GU, GV, and GI, respectively. We use constant smoothing parameters that are carefully chosen to effectively accommodate all three bands simultaneously, ensuring consistent processing and analysis of spectra across the different grating bands. However, in future studies with real CSST data, we will investigate the potential impact of allowing these parameters to evolve in response to variations in spectral resolutions. Regarding possible spectral contamination from neighbouring sources, the data reduction pipeline will identify contaminated objects and estimate the extent of the contamination. This information will help determine whether it is appropriate to measure redshifts for these objects. Additionally, based on the specific sources of contamination, we can attempt to combine redshift results from nearby sources to eliminate issues caused by line blending or contamination from neighboring galaxies, thus improving the accuracy of redshift measurements.

% If there is a deviation in that of the real spectra, we need to adjust the standard deviations of the Gaussian kernels and the size of the median filter when smoothing the spectra and estimating the continuum. We will also adjust the minimum width when identifying emission lines, as well as the initial value and limiting conditions of the emission line shapes when Gaussian fitting.
% If the SNRs of the real spectra are different, we will adjust the threshold of emission line identification and the minimal prominence of emission line identification. If the SNR will be improved, we expect that the measurement accuracy of the emission line fluxes will also be improved, so we can strengthen the loose constraint of the emission line flux ratios in Table~\ref{ZWARNING_table}.
% Our redshift measurement using the simulated slitless spectra is without source selection in advance. If high SNR ELGs are selected in advance in the real sky survey, these spectra may show more obvious emission lines. We will adjust the number of emission lines selected for redshift measurement according to Table~\ref{emission_line_templates}.

\subsection{Main statistical results of redshift measurements}\label{sec:redshift_results}
We have conducted redshift measurements for nearly one million slitless spectra, which were randomly simulated by the CESS as previously described. A synthesis of the quality metrics for these measurements is presented in Table~\ref{tab:redshift_quality}. From this table, we observe that approximately 22.32 per cent of the total simulated galaxies have detectable redshift results, indicating that 77.68 per cent of the galaxies do not possess emission lines bright enough for detection. It is worthy to highlight that (1) definition of ELGs: galaxies displaying at least 2 emission lines with SNRs exceeding 3 and their FWHMs no less than 15 {\AA} are categorised as ELGs (this FWHM cut is adaptive to the spectral resolution and is designed to exclude false emission lines, which can manifest as noise in the spectra, thereby enhancing the detection of true ELGs.); (2) successful redshift criteria: a redshift measurement is deemed successful when $\left| \Delta z_\mathrm{norm} \right| < 0.01$; (3) good redshift quality: redshifts flagged with ZWARNING $=0$ are indicative of high-quality measurements. It is important to emphasize that the redshift determination from a single emission line without additional constraints is impractical for our method of measuring redshift through emission line matching. Our random sample revealed that approximately 14.53 per cent of the galaxies are classified as ELGs. The proportion of ELGs may be underestimated, because some emission lines may have FWHMs less than 15 {\AA} due to the spectral uncertainties. However, we require a stringent sample to reliably assess the quality of redshift measurements. Generally, the purity of redshift measurements reaches 85.2 per cent, increasing to 92.5 per cent for measurements tagged with ZWARNING$=0$. The overall completeness for ELGs is exceeding 95 per cent, with a over 99 per cent purity achieved for ELGs with ZWARNING$=0$.

\begin{table}
	\centering
	\caption{Qualities of redshift measurements based on emission lines for slitless spectra}
	\label{tab:redshift_quality}
	\begin{tabular}{  m{4.4cm}<{\raggedleft} | m{3cm}<{\centering}}
		\hline
        Item & Value \\\hline
        Number of spectra & 999,972 \\  %[0.25cm]
        Measurable spectra & 22.32\% \\ %[0.3cm]
        Total successful measurements & 19.01\% \\ %[0.3cm]
        Purity & 85.19\% \\%[0.25cm]
        $\sigma_\mathrm{NMAD}$ & 0.000698 \\ %[0.25cm] \\
        $median(\Delta z_\mathrm{norm})$  & $-0.000127$ \\  %[0.25cm]
        Purity with ZWARNING$=0$ & 92.54\% \\\hline %[0.3cm]
        Number of ELGs & 14.53\% \\ %[0.25cm]
        Measurable ELGs & 97.75\% \\ %[0.25cm]
        Completeness of ELGs & 95.23\% \\ %[0.25cm]
        $\sigma_\mathrm{NMAD}$ of ELGs & 0.000599 \\ %[0.25cm]
        $median(\Delta z_\mathrm{norm})$ of ELGs & $-0.000153$ \\  %[0.25cm]
        Purity of ELGs with ZWARNING$=0$  & 99.34\% \\
		\hline
	\end{tabular}
\end{table}
%During the testing of the redshift measurement process, we input all of the nearly $1,000,000$ spectra that contained wavelength calibration errors as described earlier, along with several other input parameters:
%
%\begin{itemize}
%  \item $emission\_line\_threshold$ : adaptive.
%  \item $rest\_lines$: $\text{C III] } \lambda1909$, $\text{[O II] } \lambda3727$, $\text{H}\gamma$, $\text{H}\beta$, $\text{[O III] } \lambda\lambda4959,5007$, and $\text{H}\alpha$
%  \item $redshift\_upperlimit$ : no upper limit.
%  \item $real\_redshift$: no real redshift input.
%\end{itemize}

In Fig.~\ref{data_properties}, the orange curve shows the distributions of redshift, $z$-band magnitude, and SNR of the spectra with successful redshift measurements. The parameter distributions of these successful cases are roughly the same as those of all spectra shown in blue curves. Compared with the distributions of all spectra, the successfully measured ones tend to have slightly lower redshifts, brighter magnitudes, and larger SNRs.

\subsection{Completeness, purity and accuracy}\label{Redshift Completeness and Accuracy of ELGs Based on Emission Lines}
\begin{figure}
	\includegraphics[width=\columnwidth]{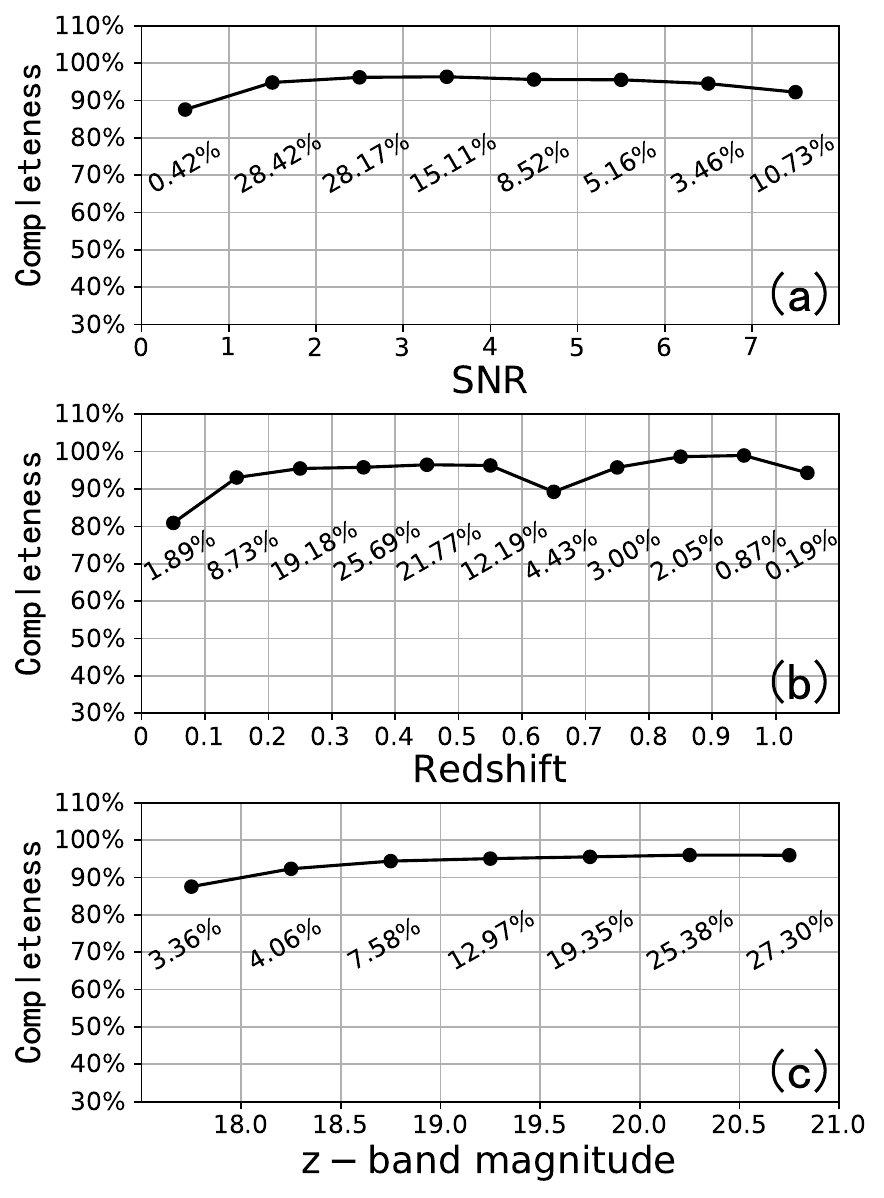}
    \caption{Completeness for ELGs as functions of spectral SNR (a), redshift (b), and $z$-band magnitude (c). The percentage in each bin represents the proportion of spectra within that bin.}
    \label{fig:completeness}
\end{figure}

\begin{figure}
	\includegraphics[width=\columnwidth]{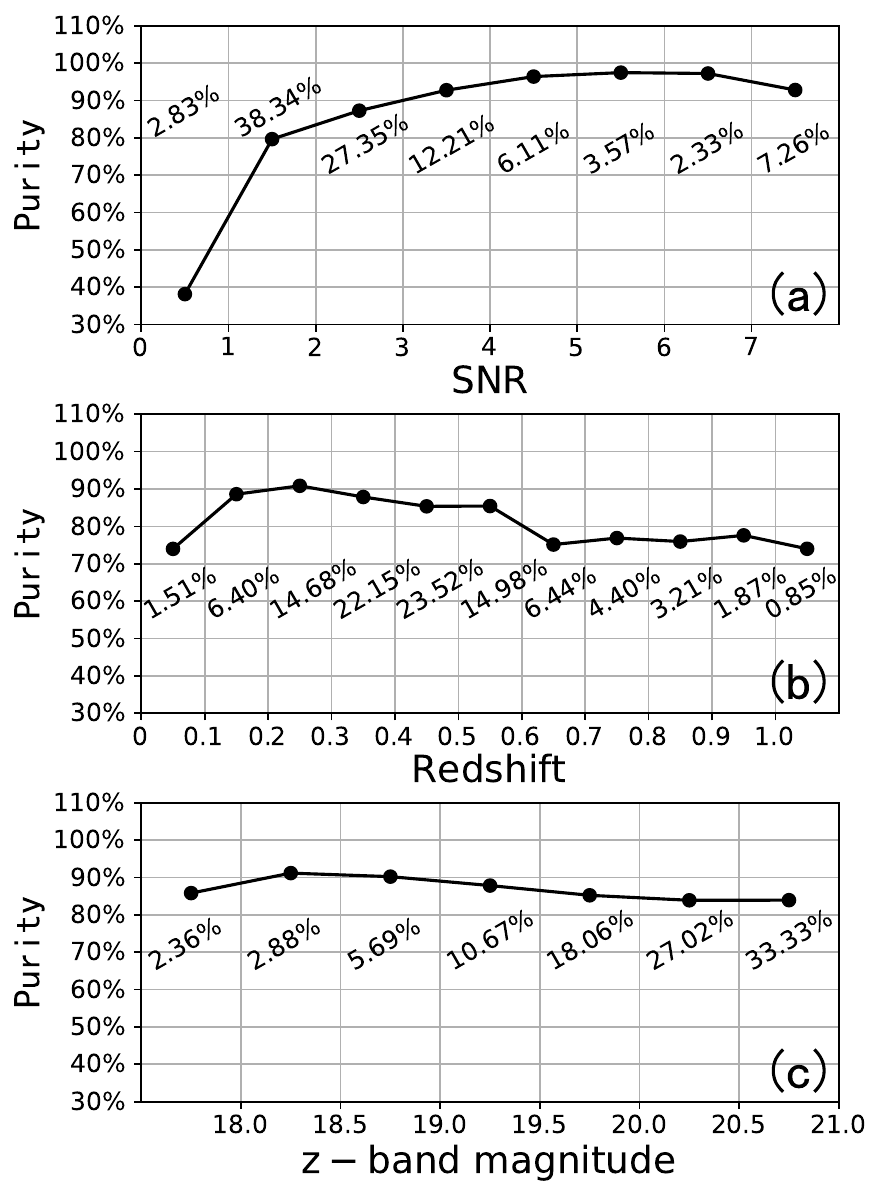}
    \caption{Purity for all measurable spectra as functions of spectral SNR (a), redshift (b), and $z$-band magnitude (c), respectively. The percentage in each bin represents the proportion of spectra within that bin.}
    \label{fig:purity}
\end{figure}

\begin{figure}
	\includegraphics[width=\columnwidth]{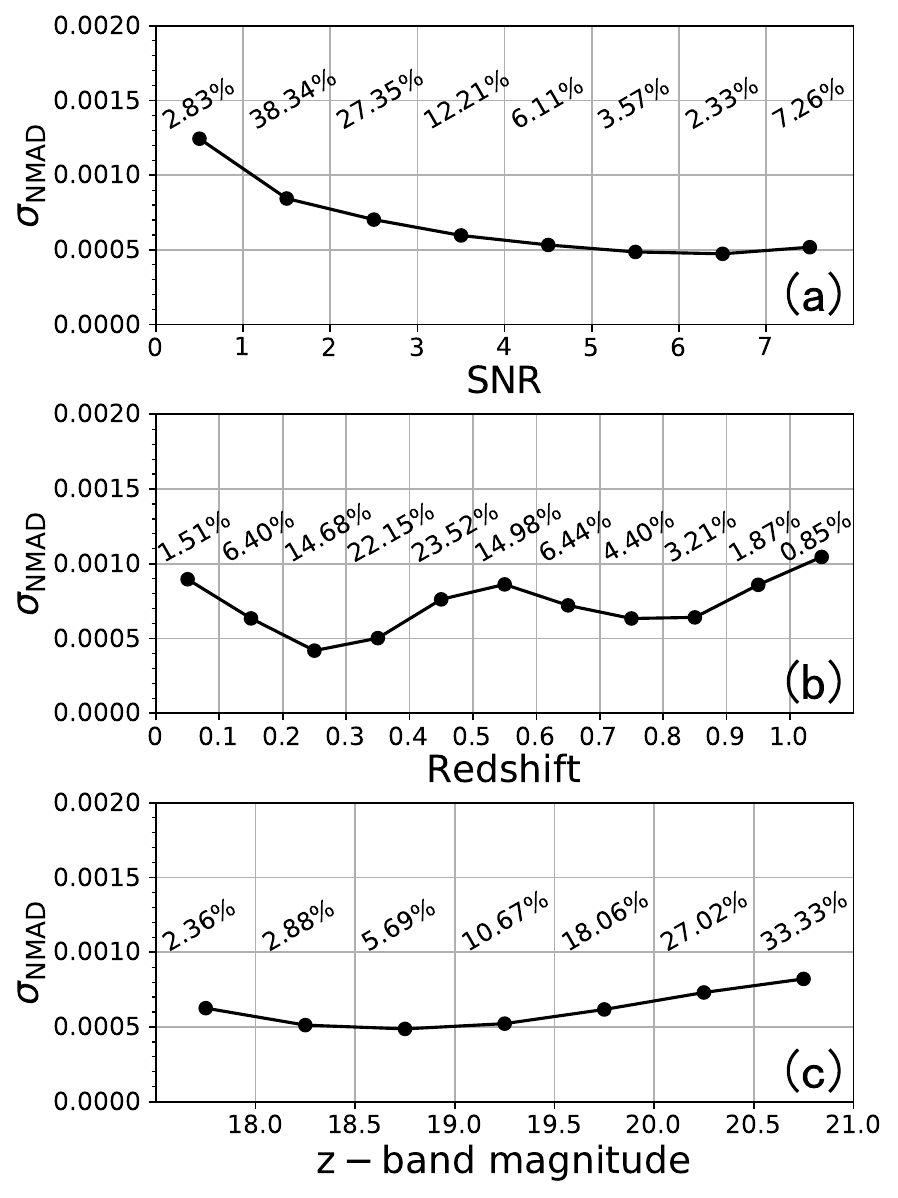}
    \caption{$\sigma_\mathrm{NMAD}$s of all successfully measured spectra as functions of spectral SNR (a), redshift (b), and $z$-band magnitude (c), respectively. The percentage in each bin represents the proportion of spectra within that bin.}
    \label{fig:accuracy}
\end{figure}

Fig.~\ref{fig:completeness} provides a comprehensive view of the redshift completeness for ELGs within our simulated slitless spectral dataset. The completeness is systematically assessed with respect to three fundamental parameters: signal-to-noise ratio (SNR), redshift, and $z$-band magnitude. The fraction of galaxies within each parameter bin is indicated next to the corresponding data point. Fig.~\ref{fig:completeness}a illustrates the variation in the completeness of redshift measurements for ELGs with respect to the SNR. As the SNR increases from its minimum value, completeness improves from 88 per cent to 96 per cent. Interestingly, the highest SNR bin shows a slight decrease in completeness, which can be attributed to the misidentification of ELGs caused by mis-estimated continua due to absorption features. A similar issue is observed in Fig.~\ref{fig:completeness}b and 7c, where galaxies in bright magnitude and low-redshift bins exhibit high spectral SNRs. This trend is also reflected in our evaluations of purity and $\sigma_\mathrm{NMAD}$, as demonstrated in Figures \ref{fig:purity} and \ref{fig:accuracy}. Fig.~\ref{fig:completeness}b depicts the completeness as a function of redshift. At most redshifts, ELG completeness exceeds 95 per cent. However, the completeness drops at $0.6<z<0.7$ due to the $\text{[O II]}\lambda3727$ emission line falling within the overlap of the GV and GI bands. Fig.~\ref{fig:completeness}c explores the relationship between the completeness and $z$-band magnitude. The completeness reaches a minimum of about 87.5 per cent for $m_z< 18.0$ and improves beyond this magnitude.
% 该失败光谱案例，SNR=202.0731138488907、红移=0.04157、MAG_Z=14.075。

\begin{figure}
	\includegraphics[width=\columnwidth]{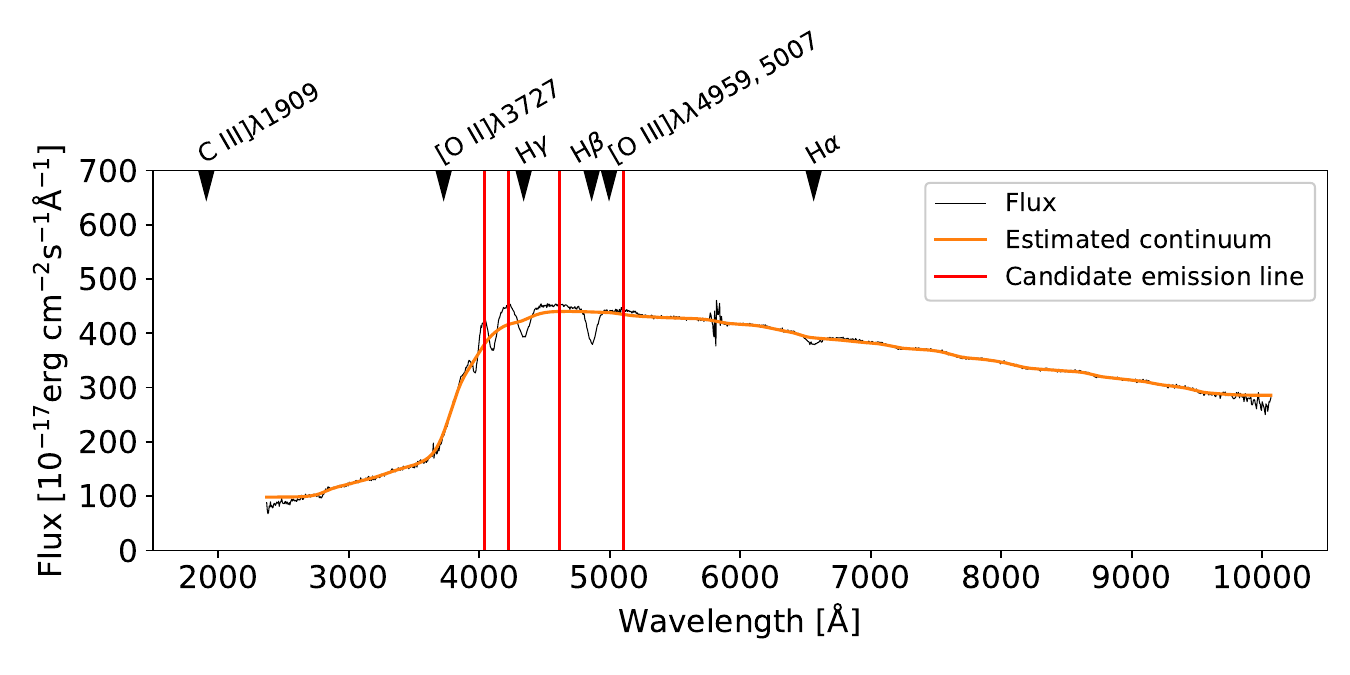}
    \caption{An example of a high-SNR spectrum at $z=0.042$ with SNR$=202$ and $m_z = 14.08$. The black curve represents the rest-frame observed spectrum, while the orange curve shows the estimated continuum. The red vertical lines indicate the detected candidate emission lines, and the black triangles mark the six template emission lines used in the analysis. In this spectrum, however, the Balmer lines are seen as absorption lines.}
    \label{fig:failed_case}
\end{figure}

In Figure \ref{fig:failed_case}, we provide an example of a high-SNR spectrum at $z=0.042$ with SNR$=202$ and $m_z = 14.08$. This spectrum displays only absorption lines, characteristics of quiescent galaxies. The estimated continuum is significantly impacted by these absorption features, leading to false peaks being misidentified as emission lines. In contrast, low-SNR spectra are dominated by noise, with the continua well represented by smoothed median-filtered data. As discussed in Section \ref{sec:em_match}, this issue can be mitigated by employing stellar population synthesis models for spectral analysis.

%\textcolor{purple}{It is because at bright magnitudes, this reduction can be ascribed to both the relative scarcity of luminous ELGs and the incorrect ELG classification. However, the proportion of bright ELGs is very small, which has little impact on the overall redshift quality.}
% Conversely, despite the abundance of objects with fainter magnitudes, their inherently lower SNR complicates redshift determinations, thereby contributing to a \textcolor{purple}{slightly} decline in completeness.

%\subsection{Redshift Purity and Accuracy Based on Emission Lines}\label{Redshift Purity and Accuracy Based on Emission Lines}

The assessment of redshift measurement purity, defined as the fraction of correct redshift determinations among all measurement results, is conducted across varying SNRs, redshifts, and $z$-band magnitudes. Fig.~\ref{fig:purity} depicts the changes in purity corresponding to these parameters. Specifically, Fig.~\ref{fig:purity}a reveals that the purity bottoms out at around 38 per cent for the lowest SNR bin, indicative of a large fraction of unsuccessful measurements. A prominent improvement is observed as SNR rises, with purity reaching about 80 per cent once SNR exceeds 1 , and over 87 per cent when SNR exceeds 2.  Fig.~\ref{fig:purity}b evidences a downturn in purity as redshift increases. Purity remains above 85 per cent at $0.1<z<0.6$ but tails off to approximately 75 per cent beyond redshift 0.6. Figure \ref{fig:purity}c shows that purity generally decreases with increasing magnitude, except in the brightest magnitude bin.

Fig.~\ref{fig:accuracy} depicts the redshift accuracy, quantified by $\sigma_\mathrm{NMAD}$, for spectra with successful redshift determinations, in relation to SNR, redshift, and $z$-band magnitude. From this figure, observe that accuracy increases with higher SNR ($\sigma_\mathrm{NMAD}$ from 0.0012 to 0.0005) and lower $z$-band magnitudes ($\sigma_\mathrm{NMAD}$ from 0.0005 to 0.0008). While the accuracy fluctuates with redshift, the impact on redshift quality is more complex due to the mixture of galaxies with different spectral SNR at the same redshift. A crucial consideration in interpreting these accuracy measures is their notable dependence on the presupposed wavelength calibration error. Specifically, a typical wavelength error of 0.1 per cent was adopted for the simulation in Paper II. 

\subsection{Effect of the number of matched lines}\label{Redshift Qualities with Matched Line Numbers}
\begin{figure}
	\includegraphics[width=\columnwidth]{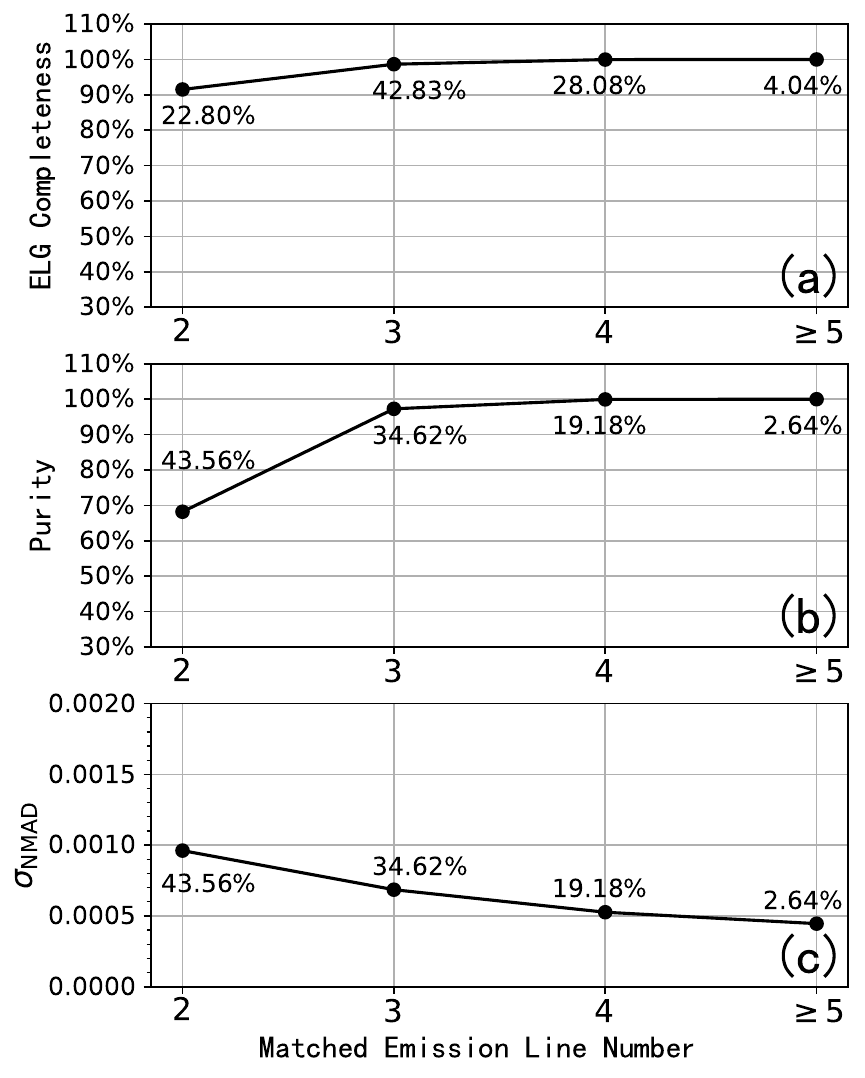}
    \caption{The ELG completeness (a), the purity (b) and accuracy $\sigma_\mathrm{NMAD}$ (c) as function of the number of matched emission lines ($N_\mathrm{line}$). The percentages next to the data points represent the proportions of spectra within those bins.}
    \label{ELG_redshift_quality_vs_matched_line_number}
\end{figure}

The quality of redshift measurements should be also related to the number of matched emission lines in the galaxy spectra. These matched emission lines encompass both the genuine emission lines inherent to the galaxy, as well as any erroneously identified lines, hence have impact on the redshift measurements. Fig.~\ref{ELG_redshift_quality_vs_matched_line_number} shows the performance metrics of completeness, purity, and $\sigma_\mathrm{NMAD}$ in relation to the number of matched emission lines ($N_\mathrm{line}$). When four or more lines are matched, the redshift measurements achieve near-perfect success rates (close to 100 per cent). With three matched lines, the figures remain high at no less than 98 per cent and 97 per cent for completeness and purity. If $N_\mathrm{line} = 2$, we can observe that a substantial dip in both completeness and purity to about 91.5 per cent and 68.2 per cent, respectively. The lower panel of Fig.~\ref{ELG_redshift_quality_vs_matched_line_number} discloses a direct inverse association between $\sigma_\mathrm{NMAD}$ of successfully measured spectra and $N_\mathrm{line}$, illustrating a decline from nearly 0.0010 at $N_\mathrm{line}=2$ down to around 0.0004 when $N_\mathrm{line} \ge 5$. 

\subsection{Redshift quality for different sample selections}
\begin{figure*}
	\includegraphics[scale=0.58]{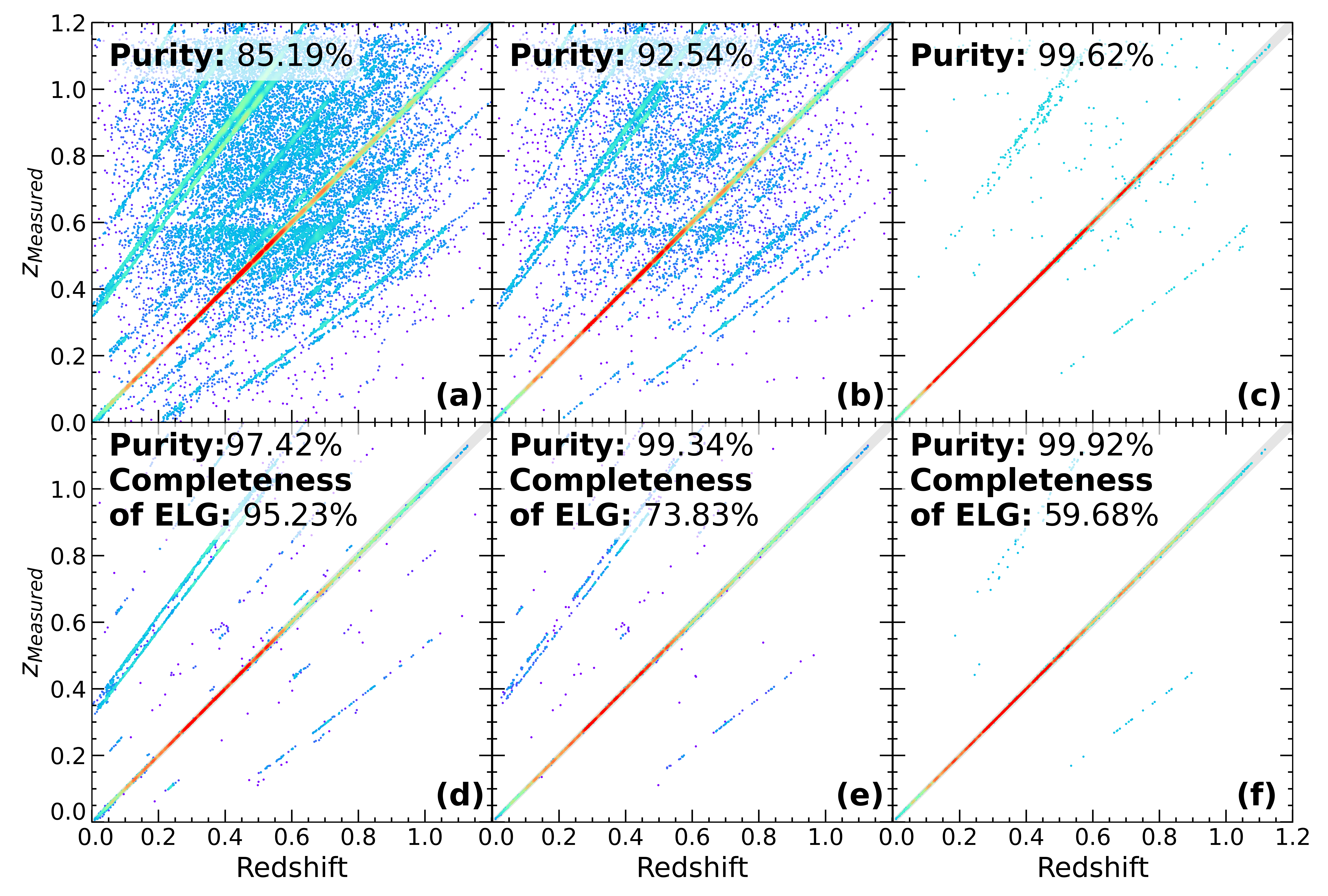}
    \caption{Comparisons of measured and true redshifts for various sample selections. The top panels correspond to measurable spectra, while the bottom panels pertain to ELGs. The left columns display comparisons for all samples. The middle columns focus on samples with ZWARNING $=0$, and the right columns showcase samples with both ZWARNING $=0$ and $N_\mathrm{line} \geq 3$. The color indicates the number density, with red representing higher density. The gray strip at each diagonal represents the allowable error range of $\left| \Delta z_\mathrm{norm} \right| < 0.01$ for measured redshifts.}
    \label{z_measured_vs_redshift}
\end{figure*}

Fig.~\ref{z_measured_vs_redshift} presents a comparison between the measured redshifts ($z_{Measured}$) with the true redshifts for different sample selections. The left columns show the comparison of all measurable spectra and identified ELGs, the middle columns display the samples where the redshift measurements have ZWARNING $=0$, and the right columns contain the samples with both ZWARNING $=0$ and $N_\mathrm{line} \geq 3$. Ideally, every measured redshift would perfectly align with its corresponding intrinsic redshift, resulting in an diagonal in each panel. The figure illustrates that points closest to the diagonals exhibit highest densities. The majority of measurable values and ELG values fall within an acceptable tolerance zone, indicating successful measurements. However, some measured redshift values notably deviate from the simulated values, suggesting unsuccessful outcomes. Interestingly, certain unsuccessful measurements cluster together, forming distinct strips. This clustering pattern may indicate systematic confusions in emission line matching, which will be discussed in Section \ref{Degeneracies in Emission Line Matching}. Overall, incorporating the redshift quality flag and the number of matched emission lines can greatly enhance the purity of the results. However, this refinement may also lead to a significant decrease in the completeness.

\section{Discussion on redshift measurements} \label{sec:discussion}

\subsection{Redshift degeneracy due to mismatching of emission lines} \label{Degeneracies in Emission Line Matching}
\begin{figure}
	\includegraphics[width=\columnwidth]{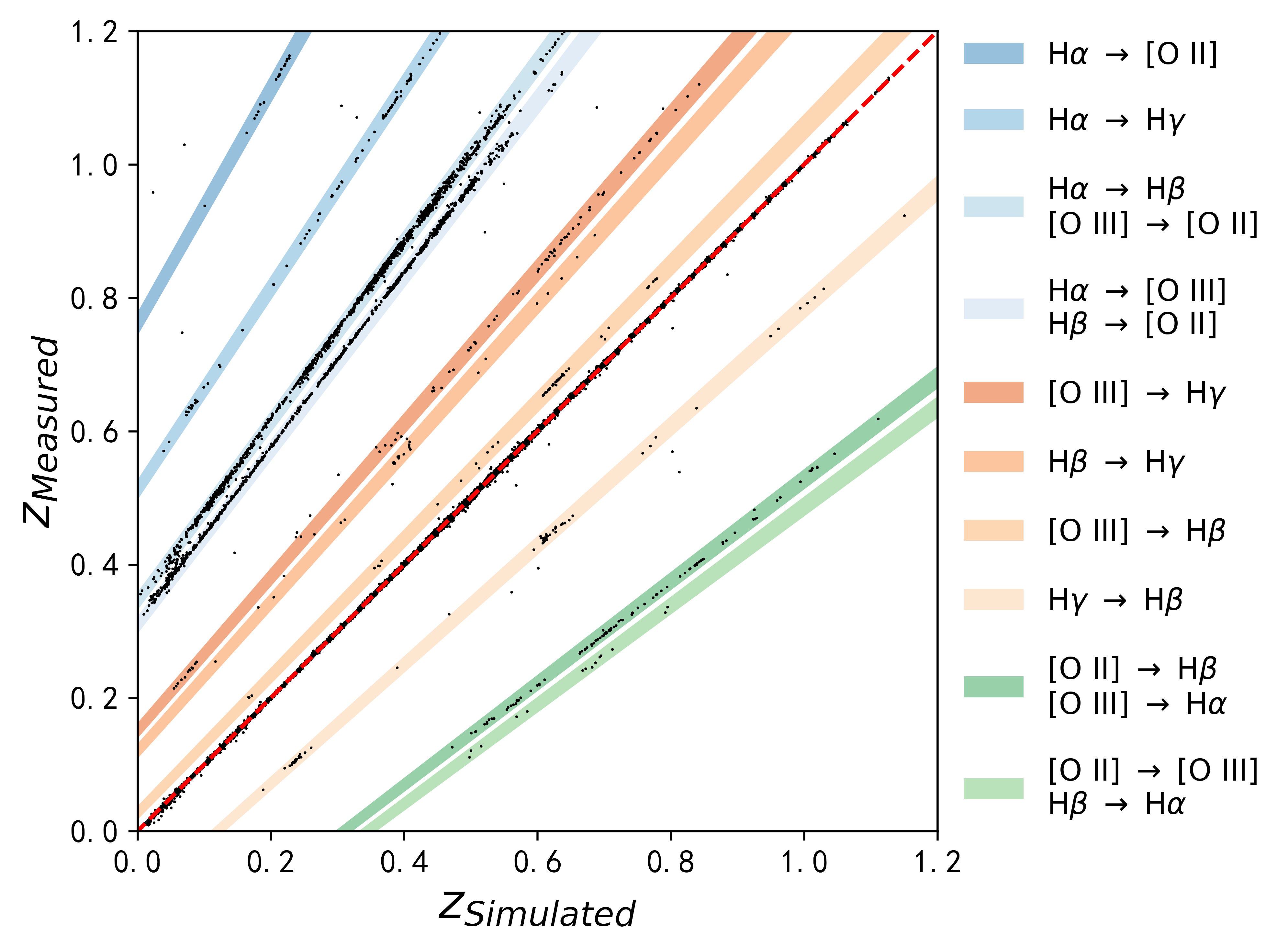}
    \caption{An example of emission line matching degeneracy of identified ELGs. The black dots correspond to the positions indicated in Fig.~\ref{z_measured_vs_redshift}(d), without the use of colour to denote density.
    The red dashed diagonal indicates the location of the successful redshift measurement. The coloured strips are the possible ranges of failed measurements caused by emission confusion, with permissible error range $\left| \Delta z_{\rm{norm}} \right| < 0.01$. }
    \label{degeneracy_cases_wr_wce_example}
\end{figure}

The erroneous redshift determinations attributed to mismatching of emission lines persist, so we employ the flux ratios of emission lines as a means to constrain our matching procedure. These mismatches can stem from inaccuracies in flux measurement due to the low SNR or the presence of non-typical physical conditions within the source, consequently giving rise to redshift ambiguities. Fig.~\ref{degeneracy_cases_wr_wce_example} illustrates the redshift degeneracy for ELGs caused by the wrong matching of emission lines. The labels in this figure aid in discerning the correlation between the measured and simulated redshifts, thereby confirming instances of wrong matched lines. The redshift degeneracy observed above the diagonal within Fig.~\ref{degeneracy_cases_wr_wce_example} arises from the misidentification of longer-wavelength emission features as shorter-wavelength ones. Conversely, beneath the diagonal, the issue lies in shorter-wavelength lines being incorrectly interpreted as longer-wavelength ones.

%The legend's $A \rightarrow B$ notation signifies that emission line A has been mistaken for line B.

Generally, two primary scenarios can lead to the incorrect matching of emission lines. Firstly, genuine lines may be misinterpreted, particularly in cases where only two detected lines are present. Due to measurement uncertainties and low spectral resolution, some cases of confusion may arise from pairs of these lines being mistaken, such as $\text{H}\alpha$ and [O III] being confused with $\text{H}\beta$ and [O II], $\text{H}\alpha$ and $\text{H}\beta$ being confused with [O III] and [O II]. Secondly, confusion can occur when detected lines are spurious. A systematic error in redshift determination can occur when one genuine line is erroneously paired with fake lines. For example, [O II] may be confused with  C III], $\text{H}\alpha$ may be confused with [O II] or $\text{H}\gamma$, $\text{H}\beta$ may be confused with $\text{H}\gamma$, and [O III] may be confused with $\text{H}\beta$. 

Table~\ref{tab:confuse} provides the fractions of spectra due to different emission-line confusions for both measurable ELGs and all measurable galaxies. The various degeneracy scenarios contribute differently to measurement failures. Typically, there is a higher incidence of situations where longer wavelength emission lines are mistakenly identified as shorter ones, resulting in systematically higher redshift measurements. This trend is primarily attributed to the fact that lower-redshift galaxies tend to exhibit relatively higher SNRs, increasing the chances of incorrect line matching. Applying a cut of ZWARNING $=0$, where the flux ratios of emission lines are confined, or requiring a minimum number of matched line can significantly reduce the occurrence of confusion cases (as shown in Fig.~\ref{z_measured_vs_redshift}).
% 描述没有confusion的比例，收到发射线confusion的比例，还有其他随机的情况。如果引入zwarning会对confusion会减少多少。
%This suggests that ZWARNING's effectiveness in resolving degeneracy differs among emission lines. When considering spectra with ZWARNING $\leq 3$, the rate of emission line confusion drops significantly, with some types of confusion nearly eliminated.

\begin{table}
	\centering
	\caption{Common Emission Line Confusions in Measurable but Failed Redshifts}
	\label{tab:confuse}
	\begin{tabular}{c | c c | c c}
		\hline
        & \multicolumn{2}{c|}{Measurable ELGs}  &   \multicolumn{2}{c}{Measurable Spectra} \\
        \hline
        & All & \makecell{ZWARNING \\ $=0$} & All & \makecell{ZWARNING \\ $=0$} \\
        \hline
        No Confusion  & $97.42\%$ & $99.34\%$ & $85.19\%$ & $92.54\%$ \\
        \hline
        \makecell{$\text{H}\alpha \rightarrow \text{H}\beta$ \\ $\text{[O III]} \rightarrow \text{[O II]}$} & $1.15\%$ & $0.26\%$ & $3.24\%$ & $1.22\%$\\
        \hline
        \makecell{$\text{H}\alpha \rightarrow \text{[O III]}$ \\ $\text{H}\beta \rightarrow \text{[O II]}$} & $1.02\%$ & $0.20\%$ & $2.55\%$ & $0.75\%$\\
        \hline
        \makecell{$\text{[O II]} \rightarrow \text{H}\beta$ \\ $\text{[O III]} \rightarrow \text{H}\alpha$} & $0.07\%$ & $0.04\%$ & $0.34\%$ & $0.15\%$\\
        \hline
        $\text{H}\alpha \rightarrow \text{H}\gamma$ & $0.06\%$ & $0.04\%$ & $0.62\%$ & $0.54\%$\\
        \hline
        $\text{H}\gamma \rightarrow \text{H}\beta$ & $0.06\%$ & $0.00\%$ & $0.16\%$ & $0.04\%$\\
        \hline
        $\text{[O III]} \rightarrow \text{H}\gamma$ & $0.05\%$ & $0.03\%$ & $0.74\%$ & $0.39\%$\\
        \hline
        $\text{[O III]} \rightarrow \text{H}\beta$ & $0.04\%$ & $0.00\%$ & $0.47\%$ & $0.11\%$\\
        \hline
        $\text{H}\alpha \rightarrow \text{[O II]}$ & $0.04\%$ & $0.04\%$ & $0.20\%$ & $0.24\%$\\
        \hline
        $\text{H}\beta \rightarrow \text{H}\gamma$ & $0.02\%$ & $0.02\%$ & $0.36\%$ & $0.20\%$\\
        \hline
        \makecell{$\text{[O II]} \rightarrow \text{[O III]}$ \\ $\text{H}\beta \rightarrow \text{H}\alpha$} & $0.01\%$ & $0.00\%$ & $0.09\%$ & $0.03\%$\\
        \hline
        others & $0.06\%$ & $0.02\%$ & $6.04\%$ & $3.78\%$ \\
		\hline
	\end{tabular}
\end{table}

\subsection{Redshift success rate of top three redshift outputs}\label{Redshift Covering Rate}
Our redshift measurement process allows for a maximum of three redshift estimates. However, confusion arising from matching emission lines can impact the precision of redshift estimation, potentially affecting the accuracy of the primary redshift estimate. As a result, the secondary and tertiary estimates, while ranked lower, may still represent plausible alternatives. A "successful redshift cover" is defined when at least one of top three redshift outputs is correct. Correspondingly, the redshift covering rate is calculated as the proportion of successful covers among the total redshift measurements.

Table~\ref{tab:cover} presents the success rates of the three redshift outputs for ELGs and all measurable spectra, along with the redshift covering rates for both sample groups. The results show that including secondary and tertiary redshift estimates improves the overall success rates of redshift measurements, with redshift covering rates exceeding 97 per cent for ELGs and approximately 90 per cent for all measurable spectra. This suggests that the secondary outputs can provide useful alternatives in cases where the primary redshift estimate is incorrect.

However, it is important to note that identifying an incorrect redshift without additional information or analysis is not possible within the scope of the current methodology. While incorporating external constraints, such as photometric redshifts or color information, may help refine redshift measurements and mitigate degeneracies, the effectiveness of this approach remains to be determined. Future work will explore how combining grism-based redshifts with complementary data can enhance the completeness and accuracy of redshift determinations, particularly for galaxies with only one identifiable emission line.

% \textcolor{purple}{In future research, we hope to further mitigate the degeneracies by combining additional information such as photometric redshift estimates or colors, which may help achieve higher redshift completeness and purity. We especially hope such methods in the future could assist in limiting the redshift range, so that even with only one identifiable emission line, accurate redshift measurements of galaxies can be determined. \sout{This indicates that even if the primary redshift estimate is inaccurate, reliable alternatives can be obtained from the secondary outputs. By incorporating additional information such as photometric redshift estimates or colors to constrain the redshift range and mitigate the degeneracies, we can achieve higher completeness and purity in our analyses. This also enables accurate redshift determinations for galaxies even with only one identifiable emission line.}}

\begin{table}
	\centering
	\caption{Redshift success rates for different redshift outputs}
	\label{tab:cover}
	\begin{tabular}{ rcc}
		\hline
         & All ELGs & Measurable Spectra \\\hline
        First redshift output & $95.23\%$ & $85.19\%$ \\
        Second redshift output & $2.11\%$ & $4.44\%$ \\
        Third redshift output & $0.08\%$ & $0.22\%$ \\
        Total covering rate & $97.41\%$ & $89.85\%$ \\
		\hline
	\end{tabular}
\end{table}

\section{Summary}\label{sec:summary}

%This method functions by initially detecting potential emission lines within the slitless spectra. It then aligns these lines with known rest-frame emission wavelengths and employs Gaussian fitting to refine the line centres. Utilizing this information, the algorithm calculates multiple potential redshifts through linear regression and ranks them based on goodness-of-fit metrics like chi-squared values and redshift quality indicators (ZWARNINGs) to ensure the selection of the most reliable redshift solution.
China is planning to launch the CSST with a large field of view in the near-Earth orbit. This telescope has both multi-wavelength imaging and slitless spectroscopic capabilities, enabling to conduct a sky survey covering an area of about 17,500 deg$^2$. The survey will provide unprecedented data for exploring the Universe, especially the dark matter and energy. The high spatial resolution in comparative to HST will also make it beneficial for studying structure of galaxies in the near field and faraway. The CSST slitless spectroscopy is characterised by the wide wavelength coverage, same sky coverage as the multi-wavelength imaging, and deep limiting magnitude. The accurate redshift determination will be crucial for exploring the large-scale structure and galaxy formation in relatively low-redshift universe. 

This paper describes the development and validation of a redshift measurement method relying on emission lines of galaxies, specifically designed for slitless spectra of the upcoming CSST. Key steps in determining a redshift using emission lines include identifying emission lines within the spectrum and matching them with their corresponding rest-frame wavelengths. Initially, potential emission lines are tentatively identified based on features like prominence, widths, and SNRs in the processed spectra. By comparing the peak wavelengths of these candidate emission lines with template emission-line wavelengths, a set of possible redshifts are derived. At each initial redshift, Gaussian fittings are employed to precisely determine the central wavelengths of emission lines, where those adjacent blended lines are fitted together. In addition, intrinsic flux ratio between different emission lines are applied to constrain the line matching and reduce the possible redshift degeneracy. Three top ranked redshift and corresponding quality flags are provided. 

The redshift measurements are validated using a sample of slitless spectra simulated by CESS, as detailed in our previous work (Paper II). Approximately one million simulated spectra are used for redshift measurement analysis. Of these, about 14.5 per cent are classified as ELGs, and the analysis shows that approximately 22.3 per cent of the total spectra yield measurable redshift results. The purity of all measurable spectra stands at 85 per cent, while the completeness for ELGs also reaches 95 per cent. The study also investigates how SNR, redshift, and magnitude impact completeness, purity, and accuracy. These metrics indicate enhancements with moderate SNR and magnitude levels alongside decreasing redshifts. Redshift degeneracies arising from the confusion of various emission line combinations are thoroughly examined, with all potential confusions enumerated. To circumvent such confusions effectively, restrictions on emission-line flux ratios are suggested. By integrating redshift quality indicators (ZWARNINGs) into the analysis, we find a notable reduction of degeneracies through stringent ZWARNING criteria. The study underscores the significance of multiple matched emission lines in enhancing completeness, purity, and the accuracy of redshift measurements. Notably, improved purity and completeness are achieved with a higher number of matched emission lines, accentuating the necessity of multiple emission line matches for robust redshift determinations. Additionally, we elucidate that even failed measurements may harbor the correct redshifts among secondary output values. However, in scenarios without ample prior information to constrain the redshift range and pinpoint the emission-line redshift, including cases with just a single emission line, we hope to make use of supplementary constraints to effectively refine the redshift determination process in the future.

The analyses presented in the study are founded solely on simulated spectra and assumptions regarding the instrument capabilities. To ensure the efficacy of emission-line redshift measurements using actual observational data from the forthcoming CSST, advancements in the configuration of the redshift measurement process are imperative, such as the parameters in spectrum smoothing and continuum estimation, \textit{threshold} in emission identification, and selection of template emission lines.

\section*{Acknowledgements}
The authors acknowledge the supports from the National Key R\&D Program of China (grant Nos. 2023YFA1607800, 2022YFA1602902, 2023YFA1607804, 2023YFA1608100, 2023YFF0714800, and 2022YFF0503400), and the National Natural Science Foundation of China (NSFC; grant Nos. 12120101003, 12373010, 12173051, and 12233008), and the Beijing Municipal Natural Science Foundation (grant No. 1222028). The authors also acknowledge the science research grants from the China Manned Space Project with Nos. CMS-CSST-2021-A02, CMS-CSST-2021-A04, CMS-CSST-2021-A05, and CMS-CSST-2021-B01 and the Strategic Priority Research Program of the Chinese Academy of Sciences with Grant Nos. XDB0550100 and XDB0550000.

\section*{Data availability}
The data involved in this article will be shared by the corresponding author upon reasonable requests.

%%%%%%%%%%%%%%%%%%%%%%%%%%%%%%%%%%%%%%%%%%%%%%%%%%
%\section*{Data Availability}

%\textcolor{red}{Data Availability}



\bibliographystyle{mnras}
\bibliography{20250217_CSST_Redshift} % if your bibtex file is called example.bib

%\begin{thebibliography}{99}
%\end{thebibliography}

%%%%%%%%%%%%%%%%%%%%%%%%%%%%%%%%%%%%%%%%%%%%%%%%%%

%%%%%%%%%%%%%%%%% APPENDICES %%%%%%%%%%%%%%%%%%%%%

%\appendix


%\textcolor{red}{Some extra material}

%%%%%%%%%%%%%%%%%%%%%%%%%%%%%%%%%%%%%%%%%%%%%%%%%%

% Don't change these lines
\bsp	% typesetting comment
\label{lastpage}
\end{document}